\documentclass[a4paper,11pt]{article}
\pdfoutput=1 % if your are submitting a pdflatex (i.e. if you have
\usepackage{jheppub} % for details on the use of the package, please
\usepackage{hyperref}
\usepackage[T1]{fontenc} % if needed
\usepackage{amsmath,amssymb}
\usepackage{slashed}
\usepackage[svgnames]{xcolor}
\usepackage{tikz}
\usepackage{mathtools}
\usepackage{upgreek}
\usepackage{subcaption}
\usepackage{todonotes}
\usetikzlibrary{arrows,positioning,cd,calc,math,decorations.pathreplacing,decorations.markings,decorations.pathmorphing%,snakes
}

\tikzset{wave/.style={decorate, decoration=snake}}
\def\nudge{.5}

\tikzset{axis/.style={ultra thick, Red!75!black, -latex, shorten <=-\nudge cm, shorten >=-2*\nudge cm}}
\tikzset{line/.style={ultra thick,black}}

\newcommand{\e}{\varepsilon}

\newcommand{\sign}{\text{sgn\,}}

\newcommand{\bs}{\boldsymbol}

\title{BPS quivers of five-dimensional SCFTs, Topological Strings and
 q-Painlev\'e equations}
\author{Giulio Bonelli,}
\author{Fabrizio Del Monte}
\author{and Alessandro Tanzini}

\affiliation{International School of Advanced Studies (SISSA),
via Bonomea 265, 34136 Trieste, Italy}
\affiliation{Institute for Geometry and Physics (IGAP),
via Beirut 2/1, 34151 Trieste, Italy}
\affiliation{INFN, Sezione di Trieste}

\abstract{
We study the discrete flows generated by the symmetry group of the 
BPS quivers for Calabi-Yau geometries describing
five dimensional superconformal quantum field theories on a circle.
These flows naturally describe the BPS particle spectrum of
such theories and at the same time generate bilinear equations of 
q-difference type which, in the rank one case, are q-Painlev\'e equations.
The solutions of these equations are shown to be given by 
grand canonical topological string partition functions which we identify with
$\tau$-functions of the cluster algebra associated to the quiver. We exemplify our construction in the case corresponding to five dimensional $SU(2)$ pure Super Yang-Mills and $N_f=2$ on a circle.
}

\begin{document} 

\maketitle
\flushbottom

\section{Introduction}

A crucial problem in quantum field theory (QFT) is
understanding its non perturbative aspects in the concrete terms of exact computations. 
QFTs can be embedded in string theory/M-theory via geometric engineering
\cite{hori2003mirror}.
Specifically, it can be obtained as the low energy limit of a compactified string theory in a large volume limit, which is needed to decouple its gravitational sector.
When QFT is obtained in this way, the nature of exact dualities gets unveiled 
through the geometric properties of the string theory background behind it: the string theory on the non-compact Calabi-Yau (CY) background geometry encodes the spectral geometry of integrable systems whose solutions 
allow to obtain exact results. 
This is possible because the non-perturbative sector of string theory, described by D-branes, gets transferred through this procedure to the geometrically engineered QFT.
The set-up engineered by M-theory compactification on $CY_3\times S^1$, in the limit of large $CY_3$ volume and finite $S^1$ radius, is that of a five dimensional supersymmetric QFT on a circle, whose particles arise from membranes 
wrapped on the 2-cycles of a suitable non compact CY manifold. 
As such, the counting of the BPS protected sectors of the theory can be 
obtained by considering a dual picture given in terms of a topological string on $CY_3$.
The precise dictionary between the two descriptions is obtained by 
identifying the topological string partition function on the $CY_3$ with 
the supersymmetric index of the gauge theory, which is conjectured to capture the exact BPS content of its 5d SCFT completion \cite{Iqbal:2012xm}.
More generally, the supersymmetric index of the gauge theory with 
surface defects is matched with the corresponding D-brane open 
topological string wave function \cite{Aganagic:2003db, Dijkgraaf:2007sw}.
The coupling constants and the moduli of the QFT arise from the geometric engineering as CY moduli parameters (K\"ahler and complex in the A and B-model picture respectively). 
Therefore, the QFT generated in this way is naturally in a generic phase in which all the coupling constants can be finite. 
To identify the weakly coupled regimes, one has to consider particular corners in the CY moduli space. In such corners, 
the topological string theory amplitudes allow a power expansion in at least one 
small parameter which is identified with the gauge coupling, while the others are fugacities of global symmetries of the QFT (masses and Coulomb parameters).

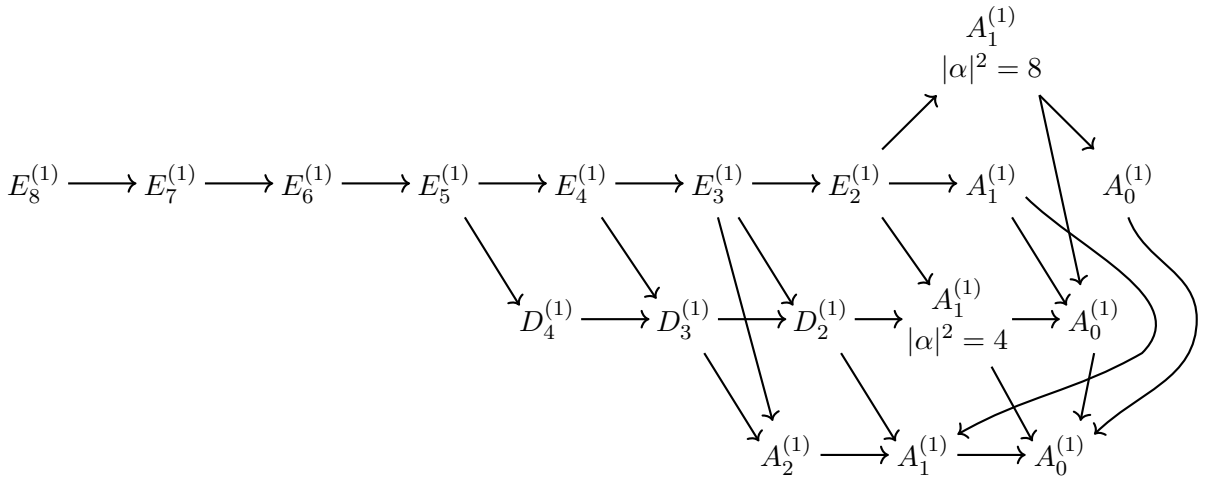
\begin{figure}[t]
\begin{center}
\begin{tikzpicture}[scale=.9]
\node at ($(0,0)$) {$E_8^{(1)}$};
\node at ($(2,0)$) {$E_7^{(1)}$};
\node at ($(4,0)$) {$E_6^{(1)}$};
\node at ($(6,0)$) {$E_5^{(1)}$}; 		%nodi in alto
\node at ($(8,0)$) {$E_4^{(1)}$};
\node at ($(10,0)$) {$E_3^{(1)}$};
\node at ($(12,0)$) {$E_2^{(1)}$};
\node at ($(14,0)$) {$A_1^{(1)}$};
\node at ($(16,0)$) {$A_0^{(1)}$};
\node at ($(14,2)$) {$\begin{matrix}A_1^{(1)} \\|\alpha|^2=8\end{matrix}$};

\draw[thick,->](0.5,0) to (1.5,0);
\draw[thick,->](2.5,0) to (3.5,0);
\draw[thick,->](4.5,0) to (5.5,0);
\draw[thick,->](6.5,0) to (7.5,0); 		%frecce orizzontali di sopra
\draw[thick,->](8.5,0) to (9.5,0);
\draw[thick,->](10.5,0) to (11.5,0);
\draw[thick,->](12.5,0) to (13.5,0);

\draw[thick,->](12.4,0.5) to (13.2,1.3);
\draw[thick,->](14.7,1.3) to (15.5,0.5);

\draw[thick,->](6.3,-0.5) to (7.1,-1.8);
\draw[thick,->](8.3,-0.5) to (9.1,-1.7);		%frecce verticali di sopra
\draw[thick,->](10.3,-0.5) to (11.1,-1.8);
\draw[thick,->](12.4,-0.5) to (13.1,-1.5);
\draw[thick,->](14.3,-0.5) to (15.1,-1.8);

\node at ($(7.5,-2)$) {$D_4^{(1)}$};
\node at ($(9.5,-2)$) {$D_3^{(1)}$};		%nodi in mezzo
\node at ($(11.5,-2)$) {$D_2^{(1)}$};
\node at ($(13.5,-2)$) {$\begin{matrix}A_1^{(1)} \\|\alpha|^2=4\end{matrix}$};
\node at ($(15.5,-2)$) {$A_0^{(1)}$};

\draw[thick,->](8,-2) to (9,-2);
\draw[thick,->](10,-2) to (11,-2);				%frecce orizzontali in mezzo
\draw[thick,->](12,-2) to (12.7,-2);
\draw[thick,->](14.3,-2) to (15,-2);

\draw[thick,->](9.8,-2.5) to (10.6,-3.8);				%frecce verticali di sotto
\draw[thick,->](11.8,-2.5) to (12.6,-3.8);
\draw[thick,->](14,-2.7) to (14.6,-3.8);

\node at ($(11,-4)$) {$A_2^{(1)}$};
\node at ($(13,-4)$) {$A_1^{(1)}$};				%nodi in basso
\node at ($(15,-4 )$) {$A_0^{(1)}$};

\draw[thick,->](11.5,-4) to (12.5,-4);				%frecce orizzontali in basso
\draw[thick,->](13.5,-4) to (14.5,-4);

\draw[thick,->] (10,-0.5) to (10.8,-3.5);
\draw[thick,->] (14.5,-0.2) to[out=315,in=45] (16.2,-2.5)  to[out=210,in=30] (13.5,-3.7) ;
\draw[thick,->] (16,-0.5) to[out=290,in=90] (17,-2) to[out=270,in=45] (15.5,-3.7);
\draw[thick,->](14.7,1.3) to (15.3,-1.5);
\draw[thick,->] (15.5,-2.5) to(15.3,-3.5);

\end{tikzpicture}
\end{center}
\caption{Sakai's Classification of discrete Painlev\'e equations by symmetry type}\label{Fig:Sakai}
\end{figure}

The problem we would like to face, in the general set-up described so far, is
that of understanding how to predict the properties of such a supersymmetric index, given the 
non-compact CY manifold which realizes the five dimensional theory via geometric engineering.
We will show that this index satisfies suitable q-difference equations which in the rank one case, namely for Calabi-Yaus whose mirror is a local genus one curve, are well-known in the mathematical physics
literature as q-Painlev\'e equations \cite{2001CMaPh.220..165S}.  
These are classified in terms of their symmetry groups as in Fig. \ref{Fig:Sakai}. Remarkably, this classification coincides with the one obtained from string theory considerations in \cite{Seiberg:1996bd}.
This allows to describe the grand canonical partition function of topological strings as $\tau$-functions of a discrete dynamical system, whose solutions
encode the BPS spectrum of the theory. From this viewpoint the grand canonical partition function is actually vector-valued in the symmetry lattice of the discrete dynamical system at hand. The exact spectrum of the relevant integrable system can be computed from the zeros of the grand partition function. 

The solutions of the discrete dynamical system are naturally parametrised in different ways according to the different BPS chambers of the theory. We will show that the Nekrasov-Okounkov \cite{Nekrasov:2003rj} presentation of the 
supersymmetric index can be recovered in the large volume regions of the Calabi-Yau moduli space which allow the geometric engineering of five-dimensional gauge theories. The expansion parameter is
schematically $e^{-V}$, $V$ being the volume of the relevant cycle corresponding to the instanton counting parameter. Around the conifold point the solution is instead naturally parameterized in terms of a matrix
model providing  the non-perturbative completion of  topological string via topological string/spectral theory correspondence \cite{Grassi:2014zfa}. The case of local $\mathbb{F}_0$ geometry, which engineers 
pure $SU(2)$ Yang-Mills in five dimensions at zero Chern-Simons level, was discussed in detail in \cite{Bonelli:2017gdk}. In this case the matrix model is a q-deformation of the $O(2)$ matrix model describing
2d Ising correlators \cite{Zamolodchikov:1994uw,Tracy:1995ax}. The quantum integrable system arising from the quantum Calabi-Yau geometry is two-particle relativistic Toda chain. 

In this paper we show that the discrete dynamics is determined from the analysis of the extended automorphism group of the BPS quiver associated to the Calabi-Yau geometry. In this respect let us recall 
the results
\cite{Cecotti:2011rv,Alim:2011ae,Cecotti:2014zga,Cirafici:2017iju}
where the BPS state spectrum of a class of four-dimensional supersymmetric theories
is generated through quiver mutations. The quiver describes the BPS vacua of the supersymmetric theory and encodes the Dirac pairing among the stable BPS particles.
The consistency of the Kontsevich-Soibelman formula \cite{Kontsevich:2008fj} for the wall crossing among the different stability chambers is encoded in Y- and Q- systems of Zamolodchikov type.
While this program have been mostly studied for four-dimensional theories, recently a proposal for BPS quivers for the five-dimensional theory on a circle has been advanced in \cite{Closset:2019juk}. 
The five-dimensional BPS quivers are conjectured to describe the BPS spectrum of the five-dimensional theory on $\mathbb{R}^4\times S^1$, and have two extra nodes with respect to the corresponding four-dimensional ones, representing, in properly chosen regimes,
the KK tower of states and the five-dimensional instanton monopole which characterise the theory on a circle.

The proposal we make in this paper is that these very same quivers also encode the q-difference equations satisfied by the SUSY index.
These are generated by studying the application of extended quiver symmetries on the relevant cluster algebra variables $\tau$, the latter being identified with 
a vector-valued topological string grand partition function. The action of the symmetry generators on the cluster algebra coefficients $y$ keeps track of the discrete flows for the $\tau$-functions. 
As such, once the gauge theory is considered on a self-dual $\Omega$-background, we obtain that its supersymmetric index satisfies a proper set of q-Painlev\'e equations generated by the extended automorphisms of the quiver.
More precisely, we identify different dynamics corresponding to different generators of the extended automorphism group. 
In a given patch, in which the topological string theory engineers a weakly coupled five dimensional theory, the generator shifting the chosen gauge coupling induces the q-Painlev\'e dynamics, while the other independent ones act as B\"acklund transformations of the former.

In this paper, we make a first step towards realizing the above proposal by 
showing that the discrete flows induced by the extended automorphism group on the BPS quiver generate in a simple way the full BPS spectrum of the 5d SCFT for some examples in the rank one case. At the same time we show that 
the Nekrasov-Okounkov dual partition function 
%\cite{Nekrasov:2009rc}
of the 5d gauge theory obtained by relevant deformation of those theories solves the  
q-Painlev\'e equations associated to the same discrete flows.
This will be accompanied also  by 
the study of the degeneration of the five-dimensional cluster algebra into the four-dimensional one by appropriate decoupling limits. More specifically,
we explore the above connection by considering in detail the case of pure $SU(2)$ gauge theory, engineered by local $\mathbb{F}_0$ and local $\mathbb{F}_1$ depending on the value of the Chern-Simons level, as well as the $SU(2)$ gauge theory with two fundamental flavors, or equivalently the one engineered by the local Calabi-Yau threefold over $dP_3$.
This case gives a much richer lattice of bilinear equations than the case of pure gauge, with four independent discrete time evolutions.

It was noticed in \cite{Bershtein:2017swf,Bershtein:2018srt,Marshakov:2019vnz} that cluster algebras provide a natural framework to describe q-deformed Painlev\'e equations, together with their higher rank generalizations and quantization (crucial to describe the refined topological string set up). Further, following the results of \cite{Bonelli:2017gdk,Bershtein:2016aef} evidence was provided for the identification of the q-Painlev\'e tau function with Topological String partition function on toric Calabi-Yau threefolds, or q-deformed conformal blocks. However, while the connection with q-Painlev\'e equations was derived in many cases, only in the case of pure $SU(N)$ gauge theory (corresponding in the $SU(2)$ case to q-Painlev\'e $\text{III}_3$) bilinear equations were derived from the cluster algebra. In this paper we derive from the cluster algebra bilinear equations for the $SU(2)$ theory with two flavors, as well as a bilinear form the the q-Painlev\'e IV equation from the cluster algebra of the local $dP_3$ geometry which to our knowledge did not appear in the literature, and we discuss its physical interpretation in terms of  the $(A_1,D_4)$ Argyres-Douglas theory.

Notice that, given the geometrical datum of the toric Calabi-Yau, it is possible to obtain its associated quiver from the corresponding dimer model \cite{Goncharov:2011hp,Yamazaki:2008bt}, and the A-cluster variables defined from this quiver lead to bilinear equations. In many cases these have been shown to be satisfied by dual partition functions of Topological String theory on this same Calabi-Yau \cite{Bershtein:2018zcz}, or by q-deformed Virasoro conformal blocks \cite{Bershtein:2016aef,jimbo2017cft,Matsuhira:2018qtx,Nagoya:2020maa}. These can also be rephrased in terms of K-theoretic blowup equations \cite{Nakajima:2005fg,Bershtein:2018zcz,Shchechkin:2020ryb}.

Here is the plan of the paper.
In Section \ref{s2} we discuss BPS quiver spectroscopy, by first 
checking our perspective with the known case of local $\mathbb{F}_0$ 
which correctly reproduces the results of \cite{Closset:2019juk}, before turning to discuss the new cases of local $\mathbb{F}_1$ and $dP_3$ five dimensional gauge theories.
In Section \ref{Sec:BPSQuivers} we study the 
discrete BPS quiver dynamics for the above examples, the related cluster algebras and obtain the explicit q-Painlev\'e equations in bilinear form.
In section \ref{sols} we show that indeed the $\tau$-functions of the
specific q-Painlev\'e flows can be matched with shifted Nekrasov-Okounkov 
partition function of the corresponding gauge theory. In Section \ref{Sec:Degeneration} we make more precise the observation that the four-dimensional quiver can be obtained by removing two nodes from the five-dimensional one, and show, for all the examples previously considered, that appropriate scaling limits of the cluster variables correctly reproduce the cluster algebra of the four-dimensional subquiver, which is known to describe the 4d BPS spectrum.
In Appendix \ref{App:Nekrasov} we collect all the necessary formulas for five-dimensional Nekrasov functions, while in Appendix \ref{Sec:Tsuda} we show how the bilinear equations of \cite{2006LMaPh..75...39T} can be also recovered from the cluster algebra approach. These are seemingly different from the ones of \cite{Matsuhira:2018qtx}, but we show that they correspond to a different choice of initial coefficients for the cluster algebra. 
Appendix \ref{saddle} collects few technical points related to improved saddle point expansion of NO partition functions used in the paper.

\section{BPS spectrum of 5d SCFT on $S^1$ and quiver mutations}\label{s2}

The construction that generates the BPS spectrum of a supersymmetric theory through mutations of its BPS quiver is known as mutation algorithm, and was widely employed in the case of four-dimensional $\mathcal{N}=2$ theories \cite{Alim:2011kw,Alim:2011ae,Cecotti:2014zga}. We recall that, given a quiver with adjacency matrix $B_{ij}$, the mutation at its $k$th node\footnote{This is an example of a cluster algebra structure, that we will introduce more thoroughly in Section \ref{Sec:BPSQuivers}.} is defined by 
\begin{equation}
\mu_k(B_{ij})= \begin{cases}
-B_{ij}, & i=k\text{ or }j=k, \\
B_{ij}+\frac{B_{ik}|B_{kj}|+B_{kj}|B_{ik}|}{2},
\end{cases}
\end{equation}
The mutations of the BPS charges $\gamma_i$ are given by
\begin{equation}\label{eq:LeftMutations}
\mu_k(\gamma_j)= \begin{cases}
-\gamma_j, & j=k, \\
\gamma_j+[B_{kj}]_+\, \gamma_k, & \text{otherwise}.
\end{cases}
\end{equation}
where we defined $[x]_+=\max(x,0) $.
In this context, each node of the quiver represents a BPS charge in the upper-half plane, and a mutation $\mu_k$ encodes the rotation of a BPS ray vector out of the upper half central charge Z-plane (see \cite{Alim:2011ae,Alim:2011kw} for a detailed description) in counterclockwise sense. If the charge is rotated out of the upper-half plane clockwise instead, one has to use a slightly different mutation rule
\begin{equation}\label{eq:RightMutations}
\widetilde{\mu_k}(\gamma_j)= \begin{cases}
-\gamma_j, & j=k, \\
\gamma_k+[-B_{kj}]_+ \,\gamma_k, & \text{otherwise}.
\end{cases}
\end{equation}
This construction is most effective when the BPS states lie in a "finite chamber", i.e. when the BPS spectrum consists entirely of hypermultiplets. This is not the case for the 5d theories we are considering: due to the intrinsically stringy origin of the UV completion of these theories, in general the BPS spectrum is organised in Regge trajectories of particles with arbitrary higher spin \cite{Galakhov:2013oja,Cordova:2015vma}; such chambers of the moduli space are known as "wild chambers". 
In \cite{Closset:2019juk} an argument was put forward for the existence of a "tame chamber" of the moduli space. 
Such a region is characterised by the fact that the higher-spin particles are unstable and decay, and one is left with hypermultiplet and vector multiplets only, giving a situation much similar to the four-dimensional weakly coupled chambers.

\subsection{Super Yang-Mills, $k=0$}

\begin{figure}
\begin{center}
\includegraphics[width=.5\textwidth]{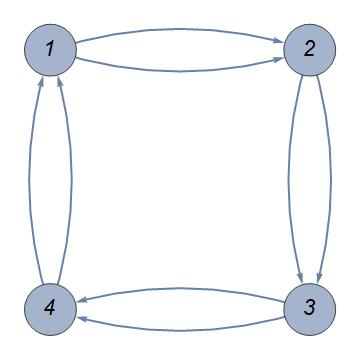}
\caption{Quiver associated to local $\mathbb{F}_0$}
\label{Fig:QuiverF0}
\end{center}
\end{figure}

\begin{figure}[h]\label{Fig:SubQuiversF0}
\begin{center}
\begin{subfigure}{.4\textwidth}
\centering
\includegraphics[width=\textwidth]{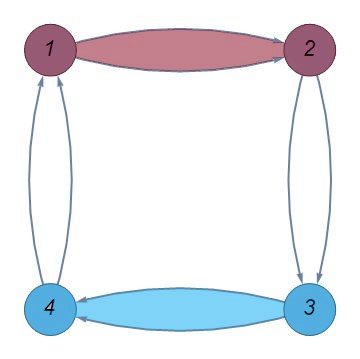}
\caption{Subquiver decomposition for $T_{\mathbb{F}_0}$}
\label{Fig:SubQuiverF0}
\end{subfigure}\hfill
\begin{subfigure}{.4\textwidth}
\centering
\includegraphics[width=\textwidth]{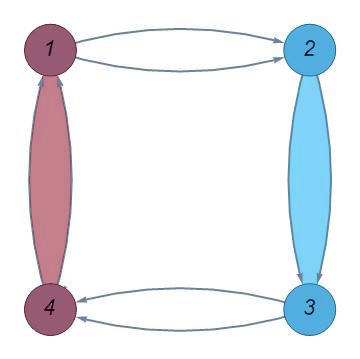}
\caption{Subquiver decomposition for $T'_{\mathbb{F}_0} $}
\label{Fig:SubQuiverF0B}
\end{subfigure}
\caption{}
\end{center}
\end{figure}

As an example, Closset and Del Zotto argued that the spectrum for local $\mathbb{F}_0$, engineering pure $SU(2)$ $SYM$ on $\mathbb{R}^4\times S^1$ with Chern-Simons level $k=0$, in such a tame chamber is organised as two copies of the weakly coupled chamber of the four-dimensional pure $SU(2)$ gauge theory. The relevant quiver is depicted in Fig.\ref{Fig:QuiverF0}, and its a
adjacency matrix is 
\begin{equation}
B=\left( \begin{array}{cccc}
0 & 2 & 0 & -2 \\
-2 & 0 & 2 & 0 \\
0 & -2 & 0 & 2 \\
2 & 0 & -2 & 0
\end{array} \right).
\end{equation}
The spectrum of this theory was originally derived by using the mutation algorithm in \cite{Closset:2019juk}. This has been done by  using the sequence of mutations
\begin{equation}
\textbf{m}=\mu_2\mu_4\mu_3\mu_1,
\end{equation}
which represents the wall-crossing arising from clockwise rotations in the upper half-plane of central charges. The n-th iteration of this operator has the following effect on the charges $\gamma_i$, $i=1,\dots,4$ of the BPS states:
\begin{align}\label{eq:WallCrossingF0}
\textbf{m}^n(\gamma_1)=\gamma_1+2n\delta_u, && \textbf{m}^n(\gamma_2)=\gamma_2-2n\delta_u, \\
\textbf{m}^n(\gamma_3)=\gamma_3+2n\delta_d, && \textbf{m}^n(\gamma_4)=\gamma_4-2n\delta_d,
\end{align}
with
\begin{align}
\delta_u=\gamma_1+\gamma_2, && \delta_d=\gamma_3+\gamma_4.
\end{align}
The action of $\textbf{m}$ corresponds to rotating out of the upper-half plane the BPS charges in the order $1342$. The towers of states obtained in this way accumulate on the vector multiplets from one side only. Because of this, the operator $\textbf{m}$ is not sufficient: in order to construct the full spectrum in this chamber, it is necessary to use also the second operator
\begin{equation}
\hat{\textbf{m}}=\hat{\mu}_1\hat{\mu}_3\hat{\mu}_2\hat{\mu}_4,
\end{equation}
constructed from right mutations \eqref{eq:RightMutations}. The shifts obtained from this operator are
\begin{align}\label{eq:WallCrossingHat}
\hat{\textbf{m}}^n(\gamma_1)=\gamma_1-2n\delta_u, && \hat{\textbf{m}}^n(\gamma_2)=\gamma_2+2n\delta_u, \\
\hat{\textbf{m}}^n(\gamma_3)= \gamma_3-2n\delta_d, && \hat{\textbf{m}}^n(\gamma_4)= \gamma_4+2n\delta_d.
\end{align}
The resulting BPS spectrum consists of two vector multiplets $\delta_u,\delta_d$, and two towers of hypermultiplets
\begin{align}
\gamma_1+n\delta_u, && \gamma_2+n\delta_u, && \gamma_3+n\delta_d, && \gamma_4+n\delta_d.
\end{align}
These are two copies of the weakly coupled spectrum of four-dimensional $\mathcal{N}=2$ $SU(2)$ pure SYM, which can be thought as being associated to the decomposition of the quiver \ref{Fig:QuiverF0} into two four-dimensional Kr\"onecker subquivers, as depicted in Figure \ref{Fig:SubQuiverF0}.

Let us show an alternative derivation of the above result, making use of the group $\mathcal{G}_Q$ of quiver automorphisms. This contains the semidirect product $Dih_4\ltimes W(A_1^{(1)})$, 
where $Dih_4$ is the dihedral group of the square, which consists only of permutations.
The automorphisms group is generated by
\begin{align}\label{eq:TimePIII3}
\pi_1=(1,3)\iota, && \pi_2=(4,3,2,1), && T_{\mathbb{F}_0}=(1,2)(3,4)\mu_1\mu_3.
\end{align}
The operator $T_{\mathbb{F}_0}$ is a Weyl translation on the $A_1^{(1)}$ lattice. 
This operator directly generates the whole BPS spectrum of the theory by acting on the charges.
Indeed, by applying the mutation rules \eqref{eq:LeftMutations} we obtain
\begin{align}
T^n_{\mathbb{F}_0}(\gamma_1)=\gamma_1+n\delta_u, && T^n_{\mathbb{F}_0}(\gamma_2)=\gamma_2-n\delta_u,\\
T^n_{\mathbb{F}_0}(\gamma_3)=\gamma_3+n\delta_d && T^n_{\mathbb{F}_0}(\gamma_4)=\gamma_4-n\delta_d.
\end{align}
By computing the Dirac pairing of these states, which is given by the adjacency matrix of the quiver, in particular
\begin{equation}
\langle\delta_u,\delta_d\rangle=\delta_u^T\cdot B\cdot\delta_d=0,
\end{equation}
we  see that $T_{\mathbb{F}_0}$ generates mutually local towers of states
\begin{align}
\gamma_1+n\delta_u, && \gamma_4+n\delta_d,
\end{align}
and
\begin{align}
\gamma_2+n\delta_u, && \gamma_3+n\delta_d.
\end{align}
As $n\rightarrow \infty$, the towers of states accumulate to the vectors $\delta_u,\delta_d$, which are vector multiplets for the four-dimensional quivers that decompose the five-dimensional quiver as in Figure \ref{Fig:SubQuiverF0}. 
The mutation operators $\textbf{m}, \hat{\textbf {m}}$  are related in a simple way to the the time evolution operator:
\begin{align}
\textbf{m}=T_{\mathbb{F}_0}^2, && \hat{\textbf{m}}= T_{\mathbb{F}_0}^{-2}\iota,
\end{align}
where $\iota$ is the inversion. From the perspective of the full automorphism group
\begin{equation}
\widetilde{W}(A_1^{(1)})\ltimes Dih_4,
\end{equation}
it is natural to consider also another translation operator
\begin{equation}
T'_{\mathbb{F}_0}=(2,3)(1,4)\mu_2\mu_4,
\end{equation}
whose action on the charges is
\begin{align}
\left(T'_{\mathbb{F}_0}\right)^{n}(\gamma_1)=\gamma_1-n(\gamma_1+\gamma_4), &&
\left(T'_{\mathbb{F}_0}\right)^{n}(\gamma_2)=\gamma_2+n(\gamma_2+\gamma_3), \\
\left(T'_{\mathbb{F}_0}\right)^{n}(\gamma_3)=\gamma_3-n(\gamma_2+\gamma_3), && 
\left(T'_{\mathbb{F}_0}\right)^{n}(\gamma_4)=\gamma_4+n(\gamma_1+\gamma_4).
\end{align}
This generates different towers of hypermultiplets, which are still organized as two copies of the weakly coupled chamber of four-dimensional super Yang-Mills, with vector multiplets
\begin{align}
\delta_l=\gamma_1+\gamma_4, && \delta_r=\gamma_2+\gamma_3.
\end{align}
In this way, we find a different infinite chamber, corresponding to the decomposition of the 5d BPS quiver as in Figure \ref{Fig:SubQuiverF0B}.

We see that considering the natural translation operators associated to the quiver automorphisms
 builds the correct spectrum for the tame chambers in a simpler way, without the need to consider both left and right mutations. This simplification occurs because we are allowing not just mutations, but also permutations, which are relabelings of the BPS charges. This operation of course has no effect on the resulting spectrum, which is the same as the one emerging from using just the mutation algorithm. However, by using quiver automorphisms, it is possible to construct more elementary dualities of the theory, and the spectrum can be constructed more simply. This plays a crucial r\^ole in more complicated cases. To illustrate this point we discuss in the following the cases of local $\mathbb{F}_1$ and $\text{dP}_3$.

\subsection{Super Yang-Mills, $k=1$}

\begin{figure}
\begin{center}
\begin{subfigure}{.4\textwidth}
\centering
\includegraphics[width=\textwidth]{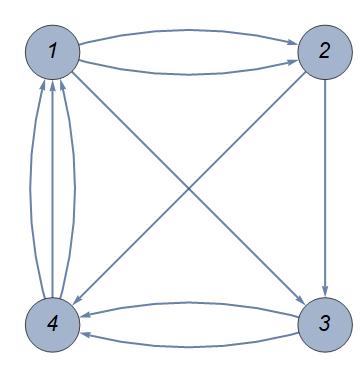}
\caption{Quiver associated to local $\mathbb{F}_1$}
\label{Fig:QuiverF1}
\end{subfigure}
\hfill
\begin{subfigure}{.4\textwidth}
\centering
\includegraphics[width=\textwidth]{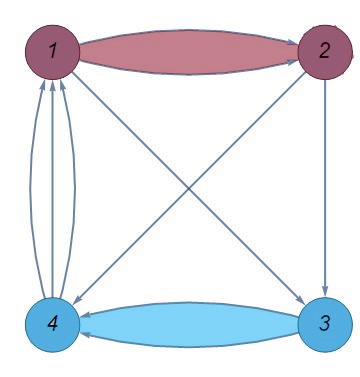}
\caption{4d subquivers for local $\mathbb{F}_1$}
\label{Fig:SubQuiverF1}
\end{subfigure}
\caption{}
\end{center}
\end{figure}

The local $\mathbb{F}_1$ quiver is displayed in Figure \ref{Fig:QuiverF1}. This engineers pure $SU(2)$ SYM with 5d Chern-Simons level $k=1$. The adjacency matrix is
\begin{equation}
B=\left( \begin{array}{cccc}
0 & 2 & 1 & -3 \\
-2 & 0 & 1 & 1 \\
-1 & -1 & 0 & 2 \\
3 & -1 & -2 & 0
\end{array} \right).
\end{equation}
The traslation operator is given by
\begin{equation}
T_{\mathbb{F}_1}=(1324)\mu_3.
\end{equation}
As far as the spectrum is concerned, this is the same as the local $\mathbb{F}_0$ one. The operators $\textbf{m},\tilde{\textbf{m}}$ are easy to build
\begin{align}
\textbf{m}\equiv T_{\mathbb{F}_1}^4=\left((1324)\mu_3 \right)^4, && \hat{\textbf{m}}\equiv T_{\mathbb{F}_1}^{-4}\iota.
\end{align}
and the associated evolution on the vector of charges $\bs{\gamma}$ is 
\begin{equation}
T_{\mathbb{F}_1}^{2n}(\bs\gamma)=T_{\mathbb{F}_0}^n(\bs\gamma)=\left( \begin{array}{cccc}
\gamma_1+n\delta_u & \gamma_2-n\delta_u & \gamma_3+n\delta_d & \gamma_4-n\delta_d
\end{array} \right),
\end{equation}
\begin{equation}
T_{\mathbb{F}_1}^{2n-1}(\bs\gamma)=\left( \begin{array}{cccc}
\gamma_3+n\delta_d, & \gamma_4-n\delta_d, & \gamma_2-n\delta_u, & \gamma_1+n\delta_u,
\end{array} \right).
\end{equation}
We see that even though the introduction of a Chern-Simons level will affect some physical aspects, it does not modify the type of states in the spectrum: again $\delta_u,\delta_d$ correspond to the vector multiplets of the 4d subquivers depicted in Figure \ref{Fig:SubQuiverF1}. What changes however is the number of tame chambers: because the symmetry group now does not include the $Dih_4$ factor -- as it is clear by inspection of the quiver -- there is not the related chamber.

\subsection{$N_f=2$, $k=0$}\label{Sec:dP3BPS}

\begin{figure}[h]
\begin{center}
\includegraphics[width=.5\textwidth]{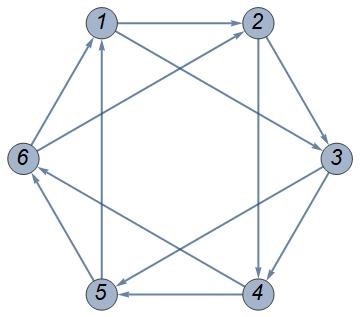}
\end{center}
\caption{Quiver for $\text{dP}_3$}
\end{figure}

\begin{figure}[h]
\begin{center}
\centering
\includegraphics[width=.5\textwidth]{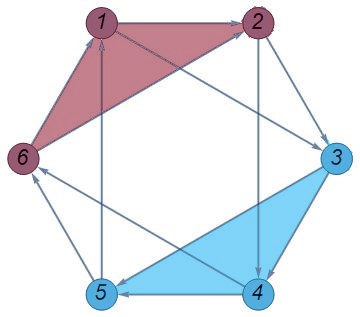}
\caption{4d subquivers for local $dP_3$, under $T_3$}
\label{Fig:SubQuiversdP33}
\end{center}
\end{figure}

When we include matter the situation is much richer, because we encounter the new feature of multiple commuting flows, each characterizing the spectrum in a different chamber of the moduli space. 
The relevant quiver is the one of $dP_3$, engineering the $SU(2)$ theory with two flavors, depicted in Figure \ref{Fig:QuiverPolydP3}. It has adjacency matrix
\begin{equation}
B=\left( \begin{array}{cccccc}
0 & 1 & 1 & 0 & -1 & -1 \\
-1 & 0 & 1 & 1 & 0 & -1 \\
-1 & -1 & 0 & 1 & 1 & 0 \\
0 & -1 & -1 & 0 & 1 & 1 \\
1 & 0 & -1 & -1 & 0 & 1 \\
1 & 1 & 0 & -1 & -1 & 0
\end{array} \right).
\end{equation}
The extended Weyl group is $\tilde{W}((A_2+A_1)^{(1)})$, which is generated by
\begin{align}
s_0=(3,6)\mu_6\mu_3 && s_1=(1,4)\mu_4\mu_1, && s_2=(2,5)\mu_5\mu_2,
\end{align}
\begin{align}
r_0=(4,6)\mu_2\mu_4\mu_6\mu_2, && r_1=(3,5)\mu_1\mu_3\mu_5\mu_1,
\end{align}
\begin{align}
\pi=(1,2,3,4,5,6), && \sigma=(1,4)(2,3)(5,6)\iota.
\end{align}
In this case there are four commuting evolution operators, given by Weyl translations of $\tilde{W}((A_2+A_1)^{(1)})$, acting on the affine root lattice $Q\left((A_2+A_1)^{(1)}\right) $ \cite{2006LMaPh..75...39T,doi:10.1063/1.4931481}\footnote{For the action on the roots, see Section \ref{Sec:dP3}.}. One has the three operators
\begin{align}
T_1=s_0s_2\pi, && T_2=s_1s_0\pi, && T_3=s_2s_1\pi
\end{align}
satisfying $T_1T_2T_3=1$, and finally 
\begin{equation}
T_4=r_0\pi^3 \, .
\end{equation}
\begin{figure}[h]
\begin{center}
\includegraphics[width=.5\textwidth]{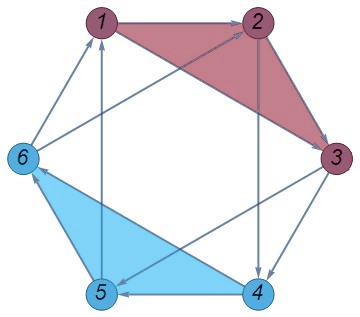}
\caption{4d subquivers for local $dP_3$, under $T_1$}
\label{Fig:SubQuiversdP31}
\end{center}
\end{figure} 
Let us consider the flow $T_1$ first, given by
\begin{equation}
T_1=s_0s_2\pi=(3,6)\mu_6\mu_3(2,5)\mu_5\mu_2(1,2,3,4,5,6).
\end{equation}
Its action on the BPS charges is the following:
%{\color{red}
%\begin{align}
%T_{1,\text{dP}_3}^n(\gamma_1)=(n+1)\gamma_1+n(\gamma_2+\gamma_3) && T_{1,\text{dP}_3}^n(\gamma_2)=\gamma_2 \\ T_{1,\text{dP}_3}^n(\gamma_3)=-(n+1)(\gamma_1+\gamma_2)-n\gamma_3 && T_{1,\text{dP}_3}^n(\gamma_4)= \gamma_4+n(\gamma_4+\gamma_5+\gamma_6) \\
% T_{1,\text{dP}_3}^n(\gamma_5)=\gamma_5 && T_{1,\text{dP}_3}^n(\gamma_6)=-(n+1)(\gamma_4+\gamma_5)-n\gamma_6 .
%\end{align} }
%{\color{blue}
	\begin{equation}
T_{1}^n: \left(\begin{array}{c}
\gamma_1 \\ \gamma_2 \\ \gamma_3 \\ \gamma_4 \\ \gamma_5 \\ \gamma_6
\end{array}\right) \longrightarrow
\left(\begin{array}{c}
(n+1)\gamma_1+n(\gamma_2+\gamma_3),\\
\gamma_2, \\ 
-(n+1)(\gamma_1+\gamma_2)-n\gamma_3,\\
\gamma_4+n(\gamma_4+\gamma_5+\gamma_6), \\
\gamma_5,\\
-(n+1)(\gamma_4+\gamma_5)-n\gamma_6
\end{array}\right)
\end{equation}
%}
We see that $T_1$ generates infinite towers of hypermultiplets given by
\begin{align}
\gamma_1+n(\gamma_1+\gamma_2+\gamma_3), && -(n+1)(\gamma_1+\gamma_2)-n\gamma_3, \\
\gamma_4+n(\gamma_4+\gamma_5+\gamma_6), && -(n+1)(\gamma_4+\gamma_5)-n\gamma_6.
\end{align}
These are the BPS states corresponding to two copies of the weakly-coupled spectrum for the $N_f=1$ theory in four dimensions, and correspond to the decomposition of the 5d quiver into two 4d subquivers for $N_f=1$, as in Figure \ref{Fig:SubQuiversdP31}. One can easily check that the towers of states are mutually local, and as $n\rightarrow\infty$ they accumulate on the rays
\begin{align}
\delta_u^{(1)}=\gamma_1+\gamma_2+\gamma_3, && \delta_d^{(1)}=\gamma_4+\gamma_5+\gamma_6,
\end{align}
which are indeed the vector multiplets for the 4d $N_f=1$ subquivers. More precisely, the towers of hypermultiplets above are only half of the towers from $N_f=1$ theory. To complete the picture here we have to consider, like in the pure gauge case, the states constructed from right mutations: these are generated as before by powers of the inverse of the evolution operator, composed with an inversion $\iota$:
\begin{equation}
T_1^{-n}\iota:\left( \begin{array}{c}
\gamma_1 \\
\gamma_2 \\
\gamma_3 \\
\gamma_4 \\
\gamma_5 \\
\gamma_6
\end{array} \right) \longrightarrow
\left(\begin{array}{c}
-n\gamma_1-(n+1)(\gamma_2+\gamma_3)  \\
\gamma_2 \\ 
n\gamma_3+(n+1)(\gamma_1+\gamma_2) \\
-n\gamma_4-(n+1)(\gamma_5+\gamma_6) \\
\gamma_5 \\
(n+1)\gamma_6+n(\gamma_5+\gamma_4).
\end{array}\right)
\end{equation}
The spectrum of the $N_f=1$ theory in four dimensions also includes two quarks, that for the subquivers in Figure \ref{Fig:SubQuiversdP31} are given by $\gamma_5,\gamma_{4}+\gamma_{6},\gamma_2,\gamma_1+\gamma_3$. We see that we recover the quarks $\gamma_5,\gamma_2$ as the states that are left invariant by $T_1$, while the other quarks would be their complementary in the subquiver. We will see below how the remaining quarks can be recovered as the states that are fixed by a different flow. 
\begin{figure}[h]
\begin{center}
\includegraphics[width=.5\textwidth]{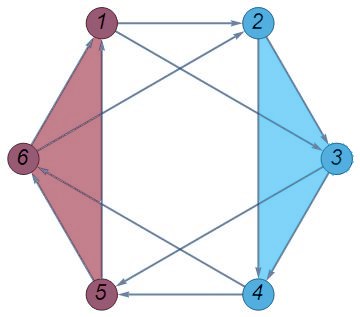}
\caption{4d subquivers for local $dP_3$, under $T_2$}
\label{Fig:SubQuiversdP32}
\end{center}
\end{figure}
$T_2$ is given by
\begin{equation}
T_2=s_1s_0\pi=(1,4)\mu_4\mu_1(3,6)\mu_6\mu_3(1,2,3,4,5,6),
\end{equation}
and acts on the BPS charges as
\begin{equation}
T_2^n:\left( \begin{array}{c}
\gamma_1 \\
\gamma_2 \\
\gamma_3 \\
\gamma_4 \\
\gamma_5 \\
\gamma_6
\end{array} \right) \longrightarrow
\left(\begin{array}{c}
-n\gamma_1-(n+1)(\gamma_5+\gamma_6) \\
(n+1)\gamma_2+n(\gamma_3+\gamma_4) \\
\gamma_3 \\
 -n(\gamma_2+\gamma_3)-n\gamma_4 \\
(n+1)\gamma_5+n(\gamma_1+\gamma_6) \\
\gamma_6.
\end{array}\right)
\end{equation}
Reasoning as before, we find that the spectrum in this chamber is organized in two copies of the 4d $N_f=1$ weakly coupled chamber, corresponding to the subquiver decomposition in Figure \ref{Fig:SubQuiversdP32}, with vector multiplets
\begin{align}
\delta_2^{(u)}=\gamma_1+\gamma_5+\gamma_6, && \delta_2^{(d)}=\gamma_2+\gamma_3+\gamma_4.
\end{align}
Finally, we have
\begin{equation}
T_3=s_2s_1\pi=(2,5)\mu_5\mu_2(1,4)\mu_4\mu_1(1,2,3,4,5,6),
\end{equation}
with action
\begin{equation}
T_3^n:\left( \begin{array}{c}
\gamma_1 \\
\gamma_2 \\
\gamma_3 \\
\gamma_4 \\
\gamma_5 \\
\gamma_6
\end{array} \right) \longrightarrow
\left(\begin{array}{c}
\gamma_1 \\
-n\gamma_2-(n+1)(\gamma_1+\gamma_6) \\
(n+1)\gamma_3+n(\gamma_4+\gamma_5) \\
 \gamma_4 \\
-n\gamma_5-(n+1)(\gamma_3+\gamma_4) \\
(n+1)\gamma_6+n(\gamma_1+\gamma_2)
\end{array}\right),
\end{equation}
which gives another chamber organized as two copies of four-dimensional weakly coupled $N_f=1$ depicted in Figure \ref{Fig:SubQuiversdP33}, with vector multiplets
\begin{align}
\delta_3^{(u)}=\gamma_1+\gamma_2+\gamma_6, && \delta_3^{(d)}=\gamma_3+\gamma_4+\gamma_5.
\end{align}
Before considering the evolution $T_4$, let us make a remark. The picture above suggests that there exists a relation between the different flows in terms of permutations of the nodes of the quiver. Indeed, it is possible to check that we have the relations
\begin{equation}\label{eq:TimesPermutations}
T_1=(3,4,5,6,1,2)T_2(5,6,1,2,3,4)=(6,1,2,3,4,5)T_3(2,3,4,5,6,1)\, .
\end{equation}
From the point of view of the BPS spectrum, it is now clear that these three flows will generate the same spectrum up to relabeling of  states, i.e. they will differ in what we call electric or magnetic in the field theory. Another interesting quiver automorphism is given by
\begin{equation}
R_2=\pi^2s_1=(3,5,1)(4,2,6)(1,4)\mu_4\mu_1,	
\end{equation}	
which is known as half-translation, because it satisfies $R_2^2=T_2$ (half-translations for $T_1,T_3$ can be obtained by using equation \eqref{eq:TimesPermutations}). Under this quiver automorphism, the BPS charges transform as
\begin{equation}
R_2^{2n+1}:\left( \begin{array}{c}
\gamma_1 \\
\gamma_2 \\
\gamma_3 \\
\gamma_4 \\
\gamma_5 \\
\gamma_6
\end{array} \right) \longrightarrow
\left(\begin{array}{c}
-(n+1)\gamma_2-n(\gamma_3+\gamma_4) \\
n\gamma_1+(n+1)(\gamma_5+\gamma_6) \\
\gamma_1+\gamma_5 \\
-(n+1)\gamma_5-n(\gamma_1+\gamma_6) \\
n\gamma_4+(n+1)(\gamma_2+\gamma_3)\\
\gamma_2+\gamma_4
\end{array}\right),
\end{equation}
while of course $R_2^{2n}=T_2^n$. Note that this generates the CPT conjugates of the towers of states as $T_2$, while the states that are left fixed by the action of $R_2$ are exactly the missing quarks from our analysis of $T_2$, so that $R_2$ generates the full spectrum of the two copies of $N_f=1$ in the subquivers of Figure \ref{Fig:SubQuiversdP32}.

Finally, the time evolution $T_4$ is given by
\begin{equation}
T_4=r_0\pi^3=(4,6)\mu_2\mu_4\mu_6\mu_2(4,5,6,1,2,3),
\end{equation}
and acts on the BPS charges as follows:
\begin{equation}
T_4^{3n-2}:\left( \begin{array}{c}
\gamma_1 \\
\gamma_2 \\
\gamma_3 \\
\gamma_4 \\
\gamma_5 \\
\gamma_6
\end{array} \right) \longrightarrow
\left(\begin{array}{c}
\gamma_1+(\gamma_3+\gamma_4)+n\delta \\
\gamma_2-(\gamma_1+\gamma_2)-n\delta \\
\gamma_3+(\gamma_5+\gamma_6)+n\delta \\
\gamma_4-(\gamma_3+\gamma_4)-n\delta \\
\gamma_5+(\gamma_1+\gamma_2)+ n\delta \\
\gamma_6-(\gamma_5+\gamma_6)-n\delta
\end{array}\right),
\end{equation}
\begin{equation}
T_4^{3n-1}:\left( \begin{array}{c}
\gamma_1 \\
\gamma_2 \\
\gamma_3 \\
\gamma_4 \\
\gamma_5 \\
\gamma_6
\end{array} \right) \longrightarrow
\left(\begin{array}{c}
\gamma_1+(\gamma_3+\gamma_4+\gamma_5+\gamma_6)+n\delta \\
\gamma_2-(\gamma_1+\gamma_2+\gamma_3+\gamma_4)-n\delta \\
\gamma_3+(\gamma_1+\gamma_2+\gamma_5+\gamma_6)+n\delta \\
\gamma_4-(\gamma_3+\gamma_4+\gamma_5+\gamma_6)-n\delta \\
\gamma_5+(\gamma_1+\gamma_2+\gamma_3+\gamma_4)+n\delta \\
\gamma_6-(\gamma_1+\gamma_2+\gamma_5+\gamma_6)-n\delta
\end{array}\right),
\end{equation}
\begin{equation}
T_4^{3n}:\left( \begin{array}{c}
\gamma_1 \\
\gamma_2 \\
\gamma_3 \\
\gamma_4 \\
\gamma_5 \\
\gamma_6
\end{array} \right) \longrightarrow
\left(\begin{array}{c}
\gamma_1+n\delta \\
\gamma_2-n\delta \\
\gamma_3+n\delta \\
\gamma_4-n\delta \\
\gamma_5+n\delta \\
\gamma_6-n\delta
\end{array}\right).
\end{equation}
We can recognise in this chamber towers of states that accumulate to the same BPS ray $\delta=\gamma_1+\dots+\gamma_6$, representing a multiplet with higher spin $s\ge1$.
In this case the four-dimensional interpretation is subtler and more interesting, and we postpone it to Section \ref{dp3}.

\section{Discrete BPS quiver dynamics, cluster algebras and q-Painlev\'e equations}\label{Sec:BPSQuivers}
The translation operators acting on the BPS quivers we described so far can be regarded as time evolution operators for discrete dynamical systems arising from deautonomization of cluster integrable systems, naturally associated to the geometric engineering
of the corresponding five-dimensional gauge theories. This allows to bridge between the BPS quiver description and classical results in the theory of q-Painlev\'e equations, and actually inspired the reformulation of the BPS quiver analysis that we presented in the previous section.
Indeed, the quivers studied in studied in \cite{Bershtein:2017swf} are exactly the 5d BPS quivers studied in section \ref{Sec:BPSQuivers}. 
In this respect, the q-Painlev\'e flows describe wall-crossing of BPS states for the 4d Kaluza-Klein theory obtained by reducing the 5d gauge theory on $S^1$. 
We now turn to  the study of all the examples we considered up to now from this perspective.

\subsection{Cluster algebras and quiver mutations}

Let us first recall the notion of cluster algebra \cite{fomin2002cluster,2006math......2259F}, as well as the two types of cluster variables that will be used throughout the paper.
	The ambient field for a cluster algebra $\mathcal{A}$ is a field $\mathcal{F}$ isomorphic to the field of rational functions in $n=\text{rk } \mathcal{A}$ independent variables, with coefficients in $\mathbb{Q}\mathbb{P}$, where $\mathbb{P}$ is the tropical semifield. The tropical semifield is defined as follows: starting with the free abelian group $(\mathbb{P},\cdot)$ with usual multiplication, the operation $\oplus$ is defined in terms of a basis\footnote{We allow for the free abelian group to have dimension less than $n$, as it will typically the case for us.} $\textbf{u}$ of $\mathbb{P}$ 
	\begin{equation}
		\prod_ju_j^{a_j}\oplus\prod_ju_j^{b_j}=\prod_ju_j^{\min(a_j,b_j)}.
	\end{equation}

	The cluster algebra $\mathcal{A}$ is determined by the choice of an initial seed. This is a triple $(Q,\bs\tau,\textbf{y})$, where

	\begin{itemize}
		\item $Q$ is a quiver without loops and 2-cycles, with $n$ vertices;
		\item $\textbf{y}=(y_1,\dots,y_n)$ is an $n$-tuple of generators of the tropical semifield $(\mathbb{P},\oplus,\cdot)$ (which in general will not be independent generators, because $\dim\mathbb{P}\le n$);
		\item $\bs\tau\equiv(\tau_1,\dots,\tau_n) $ is an n-tuple of elements of $\mathcal{F}$ forming a free generating set: they are algebraically independent over $\mathbb{Q}\mathbb{P}$, and $\mathcal{F}=\mathbb{Q}\mathbb{P}(\tau_1,\dots,\tau_n)$.	
	\end{itemize}
The variables $(\bs\tau,\textbf{y})$ are called A-cluster variables. We can alternatively define the seed as $(B,\bs\tau,\textbf{y})$ in terms of the antisymmetric adjacency matrix $B$ of the quiver.

Given these objects, the cluster algebra is the $\mathbb{ZP}$-subalgebra of $\mathcal{F}$ generated recursively by applying mutations to the initial seed. A mutation $\mu_k$ is an operation defined by its action on a seed:
\begin{equation}
\mu_k(\tau_j)= \begin{cases}
\tau_j, & j\ne k, \\
\frac{y_k\prod_{i=1}^{n}\tau_i^{[B_{ik}]_+}+\prod_{i=1}^{|Q|}\tau_i^{-[B_{ik}]_+}}{\tau_k(1\oplus y_k)}, & j=k,
\end{cases}
\end{equation}
\begin{equation}
\mu_k(y_j)= \begin{cases}
y_j^{-1}, & j=k, \\	 
y_j(1\oplus y_k^{\sign B_{jk}})^{B_{jk}}, &j\ne k,
\end{cases}
\end{equation}
\begin{equation}
\mu_k(B_{ij})= \begin{cases}
-B_{ij}, & i=k\text{ or }j=k, \\
B_{ij}+\frac{B_{ik}|B_{kj}|+B_{kj}|B_{ik}|}{2},
\end{cases}
\end{equation}
where we defined $[x]_+=\max(x,0) $.
It is clear from the above expression that the coefficients $y_i$ represent an exponentiated version of the BPS charges $\gamma_i$.

An alternative set of variables are the so-called X-cluster variables $\textbf{x}=(x_1,\dots,x_n) $, taking values in $\mathcal{F}$. They are defined in terms of the A-variables as
\begin{equation}\label{eq:ClusterVar}
x_i=y_i\prod_{j=1}^n\tau_j^{B_{ji}},
\end{equation}
and their mutation rules are the same as for coefficients, but with ordinary sum instead of semifield sum:
\begin{equation}\label{eq:MutationXCluster}
\mu_k(x_j)= \begin{cases}
x_j^{-1}, & j=k, \\	 
x_j(1+ x_k^{\sign B_{jk}})^{B_{jk}}, &j\ne k.
\end{cases}
\end{equation}

The X-cluster variables can be considered as coordinates in the so-called X-cluster variety, which is endowed with a degenerate Poisson bracket, with respect to which the X-cluster variables are log-canonically conjugated:
\begin{equation}\label{eq:ClusterPoisson}
\{x_i,x_j\}=B_{ij}x_ix_j.
\end{equation}

Given a convex Newton polygon $\Delta$ with area $S$, it is possible to construct a quiver with $2S$ nodes describing a discrete integrable system in the variables $x_i$ \cite{Goncharov:2011hp,Fock:2014ifa}. Due to \eqref{eq:ClusterPoisson}, in general the Poisson bracket is degenerate, as there is a space of Casimirs equal to $\ker(B)$. For quivers arising in this way, the quantity
\begin{equation}
q\equiv\prod_i x_i
\end{equation}
is always a Casimir. The system is integrable on the level surface
\begin{equation}
q=1.
\end{equation}
The number of independent Hamiltonians is the number of internal points of the Newton polygon. The set of discrete time flows of the integrable system is the group $\mathcal{G}_Q$ of quiver automorphisms\footnote{To be more precise, the discrete time flows are given by a subgroup $\mathcal{G}_\Delta\subset\mathcal{G}_Q$, of automorphisms preserving the Hamiltonians. These are given by spider moves of the associated dimer model, that we are not introducing here. They are very specific mutation sequences, that in this work we will rather view as Weyl translations acting on an affine root lattice. For the cases that we will be concerned with in this paper the two groups coincide and we can forget about the distinction.}. We will in fact work with the extended group $\tilde{\mathcal{G}}_Q$, that extends $\mathcal{G}_Q$ by the inclusion of the inversion operator $\iota$. This operation reverses all the arrows in the quiver, and acts on the cluster variables as
\begin{align}
\iota(x_i)= x_i^{-1}, && \iota(y_i)=y_i^{-1},
\end{align}
while the variables $\bs\tau$ are invariant, consistently with the relation \eqref{eq:ClusterVar}.

In \cite{Bershtein:2017swf} it was shown that it is possible to obtain q-Painlev\'e equations by lifting the constraint $q=1$, which amounts to the deautonomization of  the system. This is no longer integrable in the Liouville sense, since the discrete Hamiltonians are no longer preserved under the discrete flows. The related equations of motion are well-known q-difference integrable equations of mathematical physics, namely q-Painlev\'e equations: the time evolution describes in this case a foliation, whose slices are different level surfaces of the original integrable system, see e.g. \cite{joshi2019discrete} for such a description of q-Painlev\'e equations. These equations can be obtained geometrically by studying configurations of blowups of eight points on $\mathbb{P}^1\times\mathbb{P}^1$, or equivalently by configurations of nine blowups on $\mathbb{P}^2$. As in the case of differential Painlev\'e equations \cite{Okamoto1979SurLF}, this leads to a classification in terms of the space of their initial conditions, called in this context surface type of the equation, or equivalently  by their symmetry groups due to Sakai \cite{2001CMaPh.220..165S}, see Figure \ref{Fig:Sakai}. The former are given by an affine algebra, while the latter turns out to be given by the extended Weyl group of another affine algebra, which is the orthogonal complement of the first one in the group of divisors $Pic(X)$, $X$ being the surface obtained by blowing up points on $\mathbb{P}^1\times\mathbb{P}^1$.

It was further argued in \cite{Bershtein:2017swf} that the time evolution given by the deautonomization of the cluster integrable system, when written in terms of the cluster A-variables $(\bs\tau,\textbf{y})$ takes the form of bilinear equations, so that we can identify the variables $\bs\tau$ with tau functions for q-Painlev\'e equations. However, while the q-Painlev\'e equations in terms of the X-cluster variables were derived for all the Newton polygons with one internal point in \cite{Bershtein:2017swf}, their bilinear form was not obtained, except for the Newton polygon of local $\mathbb{F}_0$, corresponding to the q-Painlev\'e equation of surface type $A_7^{(1)'}$, and local $\mathbb{F}_1$ in \cite{Bershtein:2018srt}, corresponding to $A_7^{(1)}$ in Sakai's classification. 

 In the next section we review these two cases, before turning to the case of $dP_3$, which corresponds instead to the surface type $A_5^{(1)}$. In fact, this case is much richer, as it admits four commuting discrete flows: we will show that one of these reproduces the bilinear equations considered in \cite{jimbo2017cft,Matsuhira:2018qtx} for q-Painlev\'e $\text{III}_1$.

\subsection{Pure gauge theory and q-Painlev\'e III$_3$}
\label{Sec:PureGauge}

Let us briefly review how q-Painlev\'e equations are obtained from the quivers associated to local $\mathbb{F}_0$ and local $\mathbb{F}_1$, whose Newton polygons are depicted in Figure (\ref{Fig:F0Polygon}) and (\ref{Fig:F1Polygon}). These correspond to the pure $SU(2)$ gauge theory with Chern-Simons level respectively $k=0,1$. 

\begin{figure}[h]
\begin{center}
\begin{subfigure}{.5\textwidth}
\centering
\begin{tikzpicture}[scale=1.7]
\begin{scope}
\clip (-1-\nudge ,-1-\nudge) rectangle (1+\nudge,1+\nudge);
\draw[line] (0,1) -- (1,0) -- (0,-1) -- (-1,0) -- (0,1);
\fill[Blue,opacity=.3] (0,1) -- (1,0) -- (0,-1) -- (-1,0) -- (0,1);
\end{scope}
\foreach \coord/\adj in {
  {(1,0)}/right,
  {(0,1)}/above,
  {(0,0)}/below,
  {(-1,0)}/left,
  {(0,-1)}/below,%
} {
  \fill \coord circle (2pt) node[\adj] {\coord};
}
\end{tikzpicture}
\caption{Newton polygon for local $\mathbb{F}_0$}
\label{Fig:F0Polygon}
\end{subfigure}%
\begin{subfigure}{.5\textwidth}
\centering
\begin{tikzpicture}[scale=1.7]
\begin{scope}
\clip (-1-\nudge ,-1-\nudge) rectangle (1+\nudge,1+\nudge);
\draw[line] (0,1) -- (1,0) -- (-1,-1) -- (-1,0) -- (0,1);
\fill[Blue,opacity=.3] (0,1) -- (1,0) -- (-1,-1) -- (-1,0) -- (0,1);
\end{scope}
\foreach \coord/\adj in {
  {(1,0)}/right,
  {(0,1)}/above,
  {(0,0)}/below,
  {(-1,0)}/left,
  {(-1,-1)}/below/left,%
} {
  \fill \coord circle (2pt) node[\adj] {\coord};
}
\end{tikzpicture}
\caption{Newton polygon for local $\mathbb{F}_1$}
\label{Fig:F1Polygon}
\end{subfigure}
\caption{}
\end{center}
\end{figure}

\paragraph{Local $\mathbb{F}_0$:}

Let us consider first the cluster algebra associated to the quiver in Figure \ref{Fig:QuiverF0}. This corresponds to local $\mathbb{F}_0$.
The group $\mathcal{G}_Q$ of quiver automorphisms contains the symmetry group of the q-Painlev\'e equation qP$\text{III}_3$ of surface type $A_7^{(1)'} $, which is the semidirect product $Dih_4\ltimes W(A_1^{(1)})$. It is generated by
\begin{align}\label{eq:TimePIII3}
\pi_1=(1,3)\iota, && \pi_2=(4,3,2,1), && T_{\mathbb{F}_0}=(1,2)(3,4)\mu_1\mu_3.
\end{align}
The operator $T_{\mathbb{F}_0}$ generates the time evolution of the corresponding q-Painlev\'e equation, and is a Weyl translation on the underlying $A_1^{(1)}$ lattice. 
From the adjacency matrix of the quiver
\begin{equation}
B=\left( \begin{array}{cccc}
0 & 2 & 0 & -2 \\
-2 & 0 & 2 & 0 \\
0 & -2 & 0 & 2 \\
2 & 0 & -2 & 0
\end{array} \right),
\end{equation}
we see that the space of Casimirs of the Poisson bracket \eqref{eq:ClusterPoisson} is two-dimensional. We take the two Casimirs to be
\begin{align}\label{eq:qPIII3Casimirs}
q=\prod_ix_i=\prod y_i , && 
t=x_2^{-1}x_4^{-1}=y_2^{-1}y_4^{-1} .
\end{align}
Therefore, the tropical semifield has two generators, that we take to be the two Casimirs $q,t$. By fixing the initial conditions for the coefficients, consistently with equation \eqref{eq:qPIII3Casimirs}, to be
\begin{equation}
\textbf{y}=((qt)^{1/2},t^{-1/2},(qt)^{1/2},t^{-1/2}),
\end{equation}
one finds that the action of $T_{\mathbb{F}_0}$ on the coefficients yields
\begin{align}
\overline{q}=q, && \overline{t}=qt,
\end{align}
while the tau variables evolve as 
\begin{align}\label{eq:TimeEvolF0}
\begin{cases}
T_{\mathbb{F}_0}(\tau_1)=\tau_2, \\
T_{\mathbb{F}_0}(\tau_2)=\frac{\tau_2^2+(qt)^{1/2}\tau_4^2}{\tau_1}, \\
T_{\mathbb{F}_0}(\tau_3)=\tau_4, \\
T_{\mathbb{F}_0}(\tau_4)=\frac{\tau_4^2+(qt)^{1/2} \tau_2^2}{\tau_3}
\end{cases},
&&
\begin{cases}
T_{\mathbb{F}_0}^{-1}(\tau_1)=\frac{\tau_1^2+t^{1/2}\tau_3^2}{\tau_2}, \\
T_{\mathbb{F}_0}^{-1}(\tau_2)=\tau_1, \\
T_{\mathbb{F}_0}^{-1}(\tau_3)=\frac{\tau_3^2+t^{1/2}\tau_1^2}{\tau_4}, \\
T_{\mathbb{F}_0}^{-1}(\tau_4)=\tau_3,
\end{cases}
\end{align}
leading to the bilinear equations\footnote{We follow the usual convention that one overline denotes a step forward in discrete time, while one underline denotes a step backwards.}
\begin{align}
\overline{\tau_1}\underline{\tau_1}=\tau_1^2+t^{1/2}\tau_3^2, && \overline{\tau_3}\underline{\tau_3}=\tau_3^2+t^{1/2}\tau_1^2.
\end{align}
The actual q-Painlev\'e equation is the equation involving the variables \textbf{x}. It takes the form of a system of two first order q-difference equations, or of a single second-order q-difference equation, in terms of log-canonically conjugated variables
\begin{align}
F\equiv x_1, && G=x_2^{-1},
\end{align}
that satisfy
\begin{equation}
\{F,G\}=2FG.
\end{equation}
Their time evolution can be studied in a completely analogous way by using the mutation rules \eqref{eq:MutationXCluster} for X-cluster variables, and leads to the q-Painlev\'e $\text{III}_3$ equation
\begin{equation}\label{qp3}
\overline{G}\underline{G}=\left(\frac{G+t}{G+1} \right)^2.
\end{equation}

\paragraph{Local $\mathbb{F}_1$:}
We now consider the A-variables associated to the local $\mathbb{F}_1$ quiver of Figure \ref{Fig:QuiverF1}, engineering pure $SU(2)$ SYM with 5d Chern-Simons level $k=1$. The adjacency matrix is
\begin{equation}
B=\left( \begin{array}{cccc}
0 & 2 & 1 & -3 \\
-2 & 0 & 1 & 1 \\
-1 & -1 & 0 & 2 \\
3 & -1 & -2 & 0
\end{array} \right).
\end{equation}
The corresponding equation is the q-Painlev\'e equation of surface type $A_7^{(1)}$, which is a different q-discretization of the differential Painlev\'e $\text{III}_3$. The time evolution is given by
\begin{equation}
T_{\mathbb{F}_1}=(1324)\mu_3.
\end{equation}
The Casimirs are now
\begin{align}\label{eq:CasF1Def}
q=\prod_iy_i, && t=y_1y_2^{-1}y_3^2.
\end{align}
Consistently with this relation, we choose the following initial conditions for the coefficients:
\begin{equation}
\textbf{y}=(t^{1/2},t^{1/2},t^{1/2},qt^{-3/2}).
\end{equation}
This yields the time evolution of the Casimirs
\begin{align}\label{eq:CasimirF1}
\overline{q}=q, && \overline{t}=q^{1/2}t,
\end{align}
and of the $\tau$-variables
\begin{align}
\begin{cases}
T_{\mathbb{F}_1}(\tau_1)=\tau_4 \\
T_{\mathbb{F}_1}(\tau_2)=\frac{\tau_4^2+t^{1/2}\tau_1\tau_2}{\tau_3}, \\
T_{\mathbb{F}_1}(\tau_3)=\tau_1, \\
T_{\mathbb{F}_1}(\tau_4)=\tau_2.
\end{cases} &&
\begin{cases}
T_{\mathbb{F}_1}^{-1}(\tau_1)=\tau_3, \\
T_{\mathbb{F}_1}^{-1}(\tau_2)=\tau_4, \\
T_{\mathbb{F}_1}^{-1}(\tau_3)=\frac{t^{1/2}\tau_1^2+\tau_3\tau_4}{\tau_2}, \\
T_{\mathbb{F}_1}^{-1}(\tau_4)=\tau_1.
\end{cases}
\end{align}
The bilinear equations obtained in this way are
\begin{align}
\overline{\tau}_4\underline{\tau_3}=t^{1/2}\tau_1^2+\tau_3\tau_4, && \overline{\tau}_2\underline{\tau_1}=\tau_4^2+t^{1/2}\tau_1\tau_2,
\end{align}
which are the same as the equations appearing in \cite{Bershtein:2018srt}
\begin{equation}\label{eq:TodaBilinear}
\tau(qt)\tau(q^{-1}t)=\tau^2+t^{1/2}\tau(q^{1/2}t)\tau(q^{-1/2}t)
\end{equation}
for the single tau function $\tau_4\equiv\tau$. The identification is achieved by noting that
\begin{equation}
\begin{cases}
\tau_1=\underline{\tau_4}, \\
\tau_2=\overline{\tau_1}, \\
\tau_3=\underline{\underline{\tau_1}},
\end{cases}
\end{equation}
so that the first of our bilinear equations becomes
\begin{equation}
\overline{\overline{\tau}}\underline{\underline{\tau}}=\tau^2+t^{1/2} \overline{\tau}\underline{\tau},
\end{equation}
which coincides with \eqref{eq:TodaBilinear} after using \eqref{eq:CasimirF1}.

\subsection{Super Yang-Mills with two flavors and qPIII$_1$}\label{Sec:dP3}

\begin{figure}[h]
\begin{center}
\begin{subfigure}{.5\textwidth}
\centering
\begin{tikzpicture}[scale=2]
\begin{scope}
\clip (-1-\nudge ,-1-\nudge) rectangle (1+\nudge,1+\nudge);
\draw[line] (0,1) -- (1,1) -- (1,0) -- (0,-1) -- (-1,-1) -- (-1,0) -- (0,1);
\fill[Blue,opacity=.3] (0,1) -- (1,1) -- (1,0) -- (0,-1) -- (-1,-1) -- (-1,0) -- (0,1);
\end{scope}
\foreach \coord/\adj in {
  {(1,0)}/right,
  {(0,1)}/above,
  {(0,0)}/below,
  {(-1,0)}/left,
  {(0,-1)}/below,%
  {(1,1)}/above right,
  {(-1,-1)}/below left,
} {
  \fill \coord circle (2pt) node[\adj] {\coord};
}
\end{tikzpicture}
\end{subfigure}%
\begin{subfigure}{.5\textwidth}
\centering
\includegraphics[width=\textwidth]{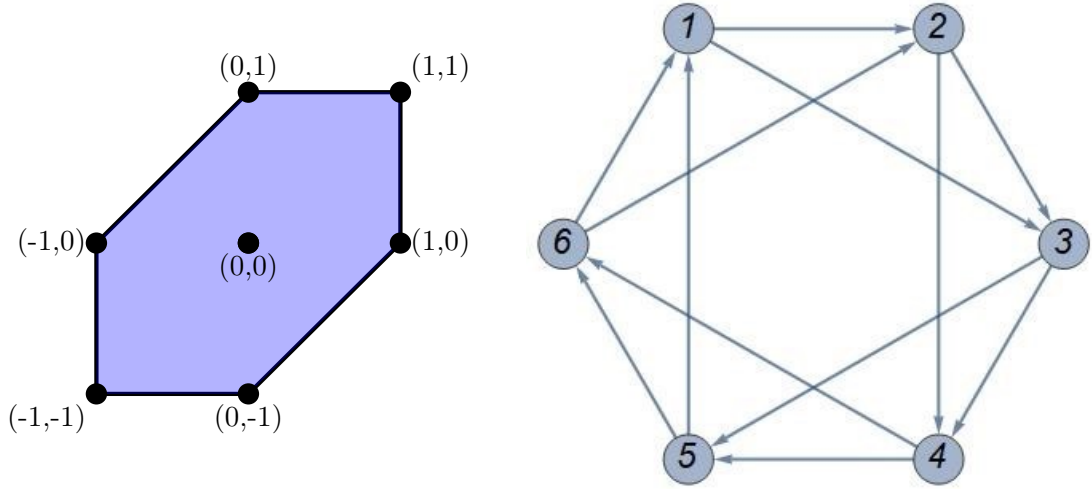}
\end{subfigure}
\end{center}
\caption{Newton polygon and quiver for $\text{dP}_3$}
\label{Fig:QuiverPolydP3}
\end{figure}

We now turn to consider the quiver associated to $dP_3$, engineering the $SU(2)$ theory with two flavors, depicted in Figure \ref{Fig:QuiverPolydP3}. It has adjacency matrix
\begin{equation}
B=\left( \begin{array}{cccccc}
0 & 1 & 1 & 0 & -1 & -1 \\
-1 & 0 & 1 & 1 & 0 & -1 \\
-1 & -1 & 0 & 1 & 1 & 0 \\
0 & -1 & -1 & 0 & 1 & 1 \\
1 & 0 & -1 & -1 & 0 & 1 \\
1 & 1 & 0 & -1 & -1 & 0
\end{array} \right),
\end{equation}
and Casimirs 
\begin{align}\label{eq:qPIIICas1}
a_0=(y_3y_6)^{-1/2}, && a_1=(y_1y_4)^{-1/2}, && a_2=(y_2y_5)^{-1/2},
\end{align}
\begin{align}\label{eq:qPIIICas2}
b_0=(y_2y_4y_6)^{-1/2}, && b_1=(y_1y_3y_5)^{-1/2},
\end{align}
that satisfy
\begin{align}
a_0a_1a_2=b_0b_1=q^{-1/2}, && q=y_1y_2y_3y_4y_5y_6.
\end{align}
As already discussed in section \ref{Sec:dP3BPS} one has in this case four commuting time evolution operators, given by Weyl translations of $\tilde{W}((A_2+A_1)^{(1)})$, which act on the affine root lattice $Q\left((A_2+A_1)^{(1)}\right) $ \cite{2006LMaPh..75...39T,doi:10.1063/1.4931481}. These time evolutions are not all associated to the same q-Painlev\'e equation: the three operators
\begin{align}
T_1=s_0s_2\pi, && T_2=s_1s_0\pi, && T_3=s_2s_1\pi
\end{align}
give rise to the q-Painlev\'e equation qP$\text{III}_1$ in the \textbf{x}-variables, and satisfy $T_1T_2T_3=1$. On the other hand, the evolution
\begin{equation}
T_4=r_0\pi^3
\end{equation}
yields q-Painlev\'e IV. 

The time evolution of the Casimirs can be obtained easily from the X-cluster variables: it is
\begin{align}\label{eq:qPIIICasimirsT1}
T_1(a_0,a_1,a_2,b_0,b_1,q)=(q^{1/2}a_0,q^{-1/2}a_1,a_2,b_0,b_1,q),
\end{align}
\begin{equation}\label{eq:qPIIICasimirsT2}
T_2(a_0,a_1,a_2,b_0,b_1,q)=(a_0,q^{1/2}a_1,q^{-1/2}a_2,b_0,b_1,q)
\end{equation}
\begin{align}\label{eq:qPIIICasimirsT3}
T_3(a_0,a_1,a_2,b_0,b_1,q)=(q^{-1/2}a_0,a_1,q^{1/2}a_2,b_0,b_1,q),
\end{align}
\begin{equation}\label{eq:qPIVCasimirsT4}
T_4(a_0,a_1,a_2,b_0,b_1,q)=(a_0,a_1,a_2,q^{-1/2}b_0,q^{1/2}b_1,q).
\end{equation}
This is the counterpart of the fact that $T_i$ are Weyl translations acting on the root lattice $Q((A_2+A_1)^{(1)}) $. If $\alpha_0,\alpha_1,\alpha_2$ are simple roots of $A_2^{(1)}$, and $\beta_0,\beta_1$ are simple roots of $A_1^{(1)}$, the action of $T_i$ as elements of the affine Weyl group is
\begin{align}
T_1(\bs\alpha,\bs\beta)=(\bs\alpha,\bs\beta)+(-1,1,0,0,0)\delta, && T_2(\bs\alpha,\bs\beta)=(\bs\alpha,\bs\beta)+(0,-1,1,0,0)\delta,
\end{align}
\begin{align}
T_3(\bs\alpha,\bs\beta)=(\bs\alpha,\bs\beta)+(1,0,-1,0,0)\delta, && T_4(\bs\alpha,\bs\beta)=(\bs\alpha,\bs\beta)+(0,0,0,1,-1)\delta,
\end{align}
where $\delta=\alpha_0+\alpha_1+\alpha_2=\beta_0+\beta_1$ is the null root of $(A_2+A_1)^{(1)} $. From each one of these discrete flows we can obtain bilinear equations for the cluster A-variables $\bs\tau$. Once we choose one of the flows as time, the other flows can be regarded as B\"acklund transformations describing symmetries of the time evolution.

Let us define the four tropical semifield generators to be $q,t,Q_1,Q_2$, and the initial condition on the parameters to be
\begin{equation}
\textbf{y}=\left(-\frac{1}{Q_2t^{1/2}},q^{1/4}t^{1/2},Q_1q^{1/4},-\frac{1}{Q_1t^{1/2}},q^{1/4}t^{1/2},Q_2q^{1/4} \right)
\end{equation}
which means, in terms of the original parametrization of the Casimirs,
\begin{align}
a_0^2=\frac{1}{Q_1Q_2q^{1/2}}, && a_1^2=Q_1Q_2t, && a_2^2=\frac{1}{q^{1/2}t},
\end{align}
\begin{align}
b_0^2=-q^{-1/2}\frac{Q_1}{Q_2}, && b_1^2=-q^{-1/2}\frac{Q_2}{Q_1}.
\end{align}

We now derive bilinear equations for the discrete flows of this geometry: the time evolution for $T_1$ is
\begin{align}
\begin{cases}
T_1(\tau_1)=\tau_3, \\
T_1(\tau_2)=\frac{\tau_5\tau_6-Q_2t^{1/2} \tau_2\tau_3}{\tau_1}, \\
T_1(\tau_4)=\tau_6, \\
T_1(\tau_5)=\frac{\tau_2\tau_3-Q_1t^{1/2} \tau_5\tau_6}{\tau_4},
\end{cases}
&&
\begin{cases}
T_1^{-1}(\tau_2)=\frac{\tau_4\tau_5+Q_1q^{1/4} \tau_1\tau_2}{\tau_3}, \\
T_1^{-1}(\tau_3)=\tau_1, \\
T_1^{-1}(\tau_5)=\frac{\tau_1\tau_2+Q_2q^{1/4}\tau_4\tau_5}{\tau_6}, \\
T_1^{-1}(\tau_6)=\tau_4.
\end{cases}
\end{align}
The action on the Casimirs is given by \eqref{eq:qPIIICasimirsT1}, that means 
\begin{align}\label{eq:T1qShifts}
T_1(Q_1)=q^{-1/2}Q_1, && T_1(Q_2)=q^{-1/2}Q_2,
\end{align}
We then have
\begin{align}
\begin{cases}
\overline{\tau_2}\underline{\tau_3}=\tau_5\tau_6-Q_2t^{1/2}\tau_2\tau_3, \\
\overline{\tau_5}\underline{\tau_6}=\tau_2\tau_3-Q_1t^{1/2}\tau_5\tau_6, \\
\underline{\tau_2}\tau_3=\tau_5\underline{\tau_6}-Q_1q^{1/2}\tau_2\underline{\tau_3}, \\
\underline{\tau_5}\tau_6=\tau_2\underline{\tau_3}-Q_2^{1/2}q^{1/2}\tau_5\underline{\tau_6},
\end{cases} &&
\overline{\tau_i}=\tau_i(q^{-1/2}Q_1,q^{1/2}Q_2).
\end{align}

The time flow under $T_2$ for the A-cluster variables is 
\begin{align}\label{eq:TimeEvoldP3}
\begin{cases}
T_2(\tau_2)=\tau_4, \\
T_2(\tau_3)=\frac{\tau_3\tau_4+q^{1/4}t^{1/2}\tau_1\tau_6}{\tau_2}, \\
T_2(\tau_5)=\tau_1, \\
T_2(\tau_6)=\frac{\tau_1\tau_6+q^{1/4}t^{1/2}\tau_3\tau_4}{\tau_5},
\end{cases}, &&
\begin{cases}
T_2^{-1}(\tau_1)=\tau_5, \\
T_2^{-1}(\tau_3)=\frac{-Q_1t^{1/2}\tau_5\tau_6+\tau_2\tau_3}{\tau_4}, \\
T_2^{-1}(\tau_4)=\tau_2, \\
T_2^{-1}(\tau_6)=\frac{-Q_2t^{1/2}\tau_2\tau_3+\tau_5\tau_6}{\tau_1},
\end{cases}
\end{align}
where the time evolution is given by
\begin{equation}
T_2(t)=qt,
\end{equation}
leading to the bilinear equations
\begin{align}
\overline{\tau_3}\tau_2=q^{1/4}t^{1/2}\overline{\tau_5}\tau_6+\tau_3\overline{\tau_2}, && \overline{\tau_6}\tau_5=\overline{\tau_5}\tau_6+q^{1/4}t^{1/2}\tau_3\overline{\tau_2}, \\
\overline{\tau_2}\underline{\tau_3}=-Q_1t^{1/2}\tau_5\tau_6+\tau_2\tau_3, && \overline{\tau_5}\underline{\tau_6}=-Q_2t^{1/2}\tau_2\tau_3+\tau_5\tau_6.
\end{align}
 In particular, from the flow $T_2$ it is possible to reproduce the bilinear equations of \cite{Matsuhira:2018qtx}, thus obtaining an explicit parametrization of the geometric quantities $a_i,b_i$ coming from the blowup configuration of $\mathbb{P}^1\times\mathbb{P}^1$ in terms of the K\"ahler parameters of $\text{dP}_3$. 

The discrete flow $T_3$ is not independent, being simply given by $T_3=T_1^{-1}T_2^{-1}$, but we write it down for completeness: 
\begin{align}
\begin{cases}
T_3(\tau_1)=\frac{\tau_1\tau_2+Q_2q^{1/4}\tau_4\tau_5}{\tau_6}, \\
T_3(\tau_3)=\tau_5, \\
T_3(\tau_4)=\frac{\tau_4\tau_5+Q_1q^{1/4} \tau_1\tau_2}{\tau_3}, \\
T_3(\tau_6)=\tau_2,
\end{cases}
&&
\begin{cases}
T_3^{-1}(\tau_1)=\frac{\tau_3\tau_4+q^{1/4}t^{1/2} \tau_1\tau_6}, \\
T_3^{-1}(\tau_2)=\tau_6, \\
T_3^{-1}(\tau_4)=\frac{\tau_1\tau_6+q^{1/4}t^{1/2} \tau_3\tau_4}{\tau_5}, \\
T_3^{-1}(\tau_5)=\tau_3,
\end{cases}
\end{align}
leading to the bilinear relations
\begin{align}
\begin{cases}
\overline{\tau_1}\underline{\tau_2}=\tau_1\tau_2+Q_2q^{1/4}\tau_4\tau_5, \\
\overline{\tau_4}\underline{\tau_5}=\tau_4\tau_5+Q_1q^{1/4}\tau_1\tau_2, \\
\underline{\tau_1}\tau_2=q^{1/4}t^{1/2}\tau_1\underline{\tau_2}+\tau_4\underline{\tau_5}, \\
\underline{\tau_4}\tau_5=q^{1/4}t^{1/2}\tau_4\underline{\tau_5}+\tau_1\underline{\tau_2}.
\end{cases},
&&
\overline{\tau_i}=\tau_i(q^{1/2}Q_1,q^{1/2}Q_2,q^{-1}t).
\end{align}

\subsection{Super Yang-Mills with two flavours, q-Painlev\'e IV and q-Painlev\'e II}\label{Sec:OtherFlowsdP3}

On top of the previous time evolutions giving rise to qPIII$_1$ equations, there is another the time evolution $T_4$ from a further automorphism of the $dP_3$ quiver. This gives rise to the qPIV dynamics and has the following action on the Casimirs, dictated by \eqref{eq:qPIVCasimirsT4}:
\begin{align}\label{eq:T4qShifts}
T_4(Q_1)=q^{1/2} Q_1, && T_4(Q_2)=q^{-1/2}Q_2.
\end{align}
On the tau variables, this amounts to
\begin{align}\label{eq:TqPIV2}
\begin{cases}
T_4(\tau_1)=\tau_4, \\
T_4(\tau_2)=\frac{\tau_1\tau_2\tau_6+\tau_4\tau_5\tau_6-Q_2t^{1/2} \tau_2\tau_3\tau_4}{\tau_1\tau_3}, \\
T_4(\tau_3)=\tau_6, \\
T_4(\tau_4)=\frac{\tau_1\tau_2\tau_6+\tau_4\tau_5\tau_6+q^{1/4}t^{1/2}\tau_2\tau_3\tau_4}{\tau_3\tau_5}, \\
T_4(\tau_5)=\tau_2, \\
T_4(\tau_6)=\frac{-Q_2t^{1/2} \tau_1\tau_2\tau_6+q^{1/4}t^{1/2} \tau_4\tau_5\tau_6-Q_2q^{1/4}t \tau_2\tau_3\tau_4}{\tau_1\tau_5},
\end{cases}
\end{align}
\begin{equation}\label{eq:TqPIV2}
\begin{cases}
T_4^{-1}(\tau_1)=\frac{\tau_1\tau_2\tau_3+Q_2q^{1/4} \tau_3\tau_4\tau_5+Q_2q^{1/2}t^{1/2} \tau_1\tau_5\tau_6}{\tau_2\tau_6}, \\
T_4^{-1}(\tau_2)=\tau_5, \\
T_4^{-1}(\tau_3)=\frac{q^{1/4}t^{1/2} \tau_1\tau_2\tau_3-Q_1t^{1/2} \tau_3\tau_4\tau_5-Q_1q^{1/4}t \tau_1\tau_5\tau_6}{\tau_2\tau_4}, \\
T_4^{-1}(\tau_4)=\tau_1, \\
T_4^{-1}(\tau_5)=\frac{\tau_1\tau_2\tau_3+Q_2q^{1/4} \tau_3\tau_4\tau_5-Q_1t^{1/2}\tau_1\tau_5\tau_6}{\tau_4\tau_6}, \\
T_4^{-1}(\tau_6)=\tau_3.
\end{cases}
\end{equation}
At first sight this seems to lead to cubic equations. However,  by following the procedure explained in Appendix \ref{app:qPIV}, one obtains an equivalent set of  bilinear equations
\begin{align}
\begin{cases}
\overline{\tau_6}\underline{\tau_2}-q^{1/4}t^{1/2} \overline{\tau_2}\underline{\tau_6}=-t^{1/2}\left(Q_2+q^{1/2}Q_1 \right) \tau_2\tau_6, \\
Q_+^{1/2}q^{1/4}\overline{\tau_6}\underline{\tau_4}+Q_2t^{1/2}\overline{\tau_4}\underline{\tau_6}=t^{1/2}\left(Q_2+q^{1/2}Q_1 \right)\tau_4\tau_6, \\
\overline{\tau_4}\underline{\tau_2}-\overline{\tau_2}\underline{\tau_4}=t^{1/2}\left(Q_2+q^{1/2}Q_1 \right)\tau_2\tau_4.
\end{cases}, &&
\overline{\tau_i}=\tau_i(q^{1/2}Q_1,q^{-1/2}Q_2).
\end{align}
These provide a bilinear form for the qPIV equation, which to our knowledge did not appear in the literature so far.

\paragraph{q-Painlev\'e II bilinear relations from "half" traslations:}

Given the root lattice $(A_2+A_1)^{(1)} $, there exists another time flow that preserves a $(A_1+A_1')^{(1)} $ sublattice only \cite{2003,2009arXiv0910.4439K}. It corresponds to the q-Painlev\'e II equation, and it is given by
\begin{equation}
R_2=\pi^2s_1,
\end{equation}
which is still an automorphism of the quiver in Figure \ref{Fig:QuiverPolydP3}. Because $R_2^2=T_2$, this flow is also known as half-translation\footnote{Other half-traslations can be analogously defined from $T_1$ and $T_3$.}.
 Its action on the Casimirs is a translational motion (i.e. a good time evolution) only on the locus $a_0=q^{-1/4} $, i.e. $Q_+=1$, on which it acts as
\begin{align}
R_2(a_1)=q^{1/4}a_1, && R_2(a_2)=q^{-1/4}a_2,
\end{align}
corresponding to
\begin{equation}
R_2(t)=q^{1/2}t.
\end{equation}
Its action on the tau-variables reads
\begin{align}
\begin{cases}
R_2(\tau_1)=\frac{\tau_3\tau_4+q^{1/4}t^{1/2} \tau_1\tau_6}{\tau_2}, \\
R_2(\tau_2)=\tau_6, \\
R_2(\tau_3)=\tau_1, \\
R_2(\tau_4)=\frac{\tau_1\tau_6+q^{1/4}t^{1/2}\tau_3\tau_4}{\tau_5}, \\
R_2(\tau_5)=\tau_3, \\
R_2(\tau_6)=\tau_4,
\end{cases} &&
\begin{cases}
R_2^{-1}(\tau_1)=\tau_3, \\
R_2^{-1}(\tau_2)=\frac{\tau_5\tau_6-Q_2t^{1/2}\tau_2\tau_3}{\tau_1}, \\
R_2^{-1}(\tau_3)=\tau_5, \\
R_2^{-1}(\tau_4)=\tau_6 , \\
R_2^{-1}(\tau_5)=\frac{\tau_2\tau_3-Q_1t^{1/2} \tau_5\tau_6}{\tau_4}, \\
R_2^{-1}(\tau_6)=\tau_2\, .
\end{cases}
\end{align}
By setting $Q_1=Q_2^{-1}\equiv Q$, we obtain the bilinear equations 
\begin{align}\label{half}
\begin{cases}
\overline{\overline{\tau_3}}\underline{\tau_6}=\tau_3\overline{\tau_6}+q^{1/4}t^{1/2}\overline{\tau_3}\tau_6, \\
\underline{\tau_3}\overline{\overline{\tau_6}}=\overline{\tau_3}\tau_6+q^{1/4}t^{1/2}\tau_3\overline{\tau_6}, \\
\overline{\tau_3}\underline{\underline{\tau_6}}=\underline{\tau_3}\tau_6-Q^{-1}t^{1/2}\tau_3\underline{\tau_6}, \\
\underline{\underline{\tau_3}}\overline{\tau_6}=\tau_3\underline{\tau_6}-Qt^{1/2}\underline{\tau_3}\tau_6,
\end{cases}
&&
\overline{\tau_i}=\tau_i(q^{1/2}t).
\end{align}
We see that in fact these equations are consistent under the further requirement $Q=-1$. This is because the third and fourth equations are obtained by simply applying $T_2^{-1}$ to the first and second one. 

According to Sakai's classification (see Figure \ref{Fig:Sakai}) and the analysis 
in \cite{Bonelli:2016qwg}
this flow correctly points to the Argyres-Douglas theory of $N_f=2$
which is, in the four dimensional limit, governed by the differential PII equation.
It would be interesting to see if the relevant $\tau$-function can be constructed from a 5d lift of the matrix model considered in \cite{Grassi:2018spf}.

\subsection{Summary of dP$_3$ bilinear equations}\label{Sec:DP3Recap}
To conclude this Section, let us collect here all the flows we found for $dP_3$ geometry together with the respective bilinear equations:
\begin{align}
\begin{array}{c}
T_1,\text{ qPIII}_1,\\
\overline{Q_+}=q^{-1}Q_+
\end{array}
, && \begin{cases}
\overline{\tau_2}\underline{\tau_3}=\tau_5 \tau_6-Q_2t^{1/2}\tau_2\tau_3, \\
\overline{\tau_5}\underline{\tau_6}=\tau_2\tau_3-Q_1t^{1/2}\tau_5\tau_6, \\
\underline{\tau_2}\tau_3=\tau_5\underline{\tau_6} -Q_+^{1/2}q^{1/2}\tau_2\underline{\tau_3}, \\
\underline{\tau_5}\tau_6=\tau_2\underline{\tau_3} -Q_+^{1/2}q^{1/2}\tau_5\underline{\tau_6}, \\
\end{cases}
\end{align}
\begin{align}
\begin{array}{c}
T_2,\text{ qPIII}_1,\\
\overline{t}=qt,
\end{array}
&&
\begin{cases}
\overline{\tau_2}\underline{\tau_3}=\tau_2\tau_3-Q_1t^{1/2}\tau_5\tau_6, \\
\overline{\tau_5}\underline{\tau_6}=\tau_5\tau_6-Q_2t^{1/2}\tau_2\tau_3, \\
\tau_2\overline{\tau_3}=\overline{\tau_2}\tau_3+q^{1/4}t^{1/2}\overline{\tau_5}\tau_6, \\
\overline{\tau_5}\tau_6=\tau_5\overline{\tau_6}+q^{1/4}t^{1/2}\overline{\tau_2}\tau_3,
\end{cases}
\end{align}
\begin{align}
\begin{array}{c}
T_3,\text{ qPIII}_1,\\
\overline{t}=t/q,\text{ }\overline{Q_+}=qQ_+
\end{array}
&&
\begin{cases}
\overline{\tau_1}\underline{\tau_2}=\tau_1\tau_2+Q_+^{1/2}q^{1/4}\tau_4\tau_5, \\
\overline{\tau_4}\underline{\tau_5}=\tau_4\tau_5+Q_+^{1/2}q^{1/4}\tau_1\tau_2, \\
\underline{\tau_1}\tau_2=q^{1/4}t^{1/2} \tau_1\underline{\tau_2}+\tau_4\underline{\tau_5}, \\
\underline{\tau_4}\tau_5=q^{1/4}t^{1/2} \tau_4\underline{\tau_5}+\tau_1\underline{\tau_2}, \\
\end{cases}
\end{align}
\begin{align}
\begin{array}{c}
T_4, \text{qPIV, } \\
\overline{Q_-}=qQ_-
\end{array}
&&
\begin{cases}
\overline{\tau_6} \underline{\tau_2}-q^{1/4}t^{1/2} \overline{\tau_2}\underline{\tau_6}+(tqQ_+)^{1/2}\left(1+Q_-^{1/2}q^{-1/2}\right)\tau_2 \tau_6=0, \\
Q_+^{1/2}q^{1/4} \overline{\tau_6} \underline{\tau_4}+q^{1/4}t^{1/2}\overline{\tau_4}\underline{\tau_6}-(tqQ_+)^{1/2}\left(1+Q_-^{1/2}q^{-1/2}\right)\tau_4 \tau_6=0, \\
\overline{\tau_4} \underline{\tau_2}  -\overline{\tau_2}\underline{\tau_4}-(tqQ_+)^{1/2}\left[1+Q_-^{1/2}q^{-1/2}\right]\tau_2 \tau_4=0,
\end{cases}
\end{align}
\begin{align}
\begin{array}{c}
R_2,\text{ qPII}, (Q_1=Q_2^{-1}=-1)\\
\overline{t}=q^{1/2}t,
\end{array}
&&
\begin{cases}
\overline{\overline{\tau_3}}\underline{\tau_6}=\tau_3\overline{\tau_6}+q^{1/4}t^{1/2}\overline{\tau_3}\tau_6, \\
\underline{\tau_3}\overline{\overline{\tau_6}}=\overline{\tau_3}\tau_6+q^{1/4}t^{1/2}\tau_3\overline{\tau_6},
\end{cases}
\end{align}

\section{Solutions}\label{sols}

In this Section we discuss how the solutions of the discrete flow of BPS quivers are naturally encoded in topological string partition functions having as a target space the toric Calabi-Yau varietes associated to the relevant Newton polygons.
The corresponding geometries are given by rank two vector bundles over punctured Riemann surfaces. 
Let us recall that the BPS states of the theory are associated to curves on this geometry that locally minimise the string tension. More specifically, hypermultiplets are associated to open curves ending on the branch points of the 
covering describing the Riemann surface, while BPS vector multiplets are associated to closed curves\footnote{Let us notice that in the 5d theories 
on a circle, one finds in general also "wild chambers" with multiplets of higher spin which can be reached via wall crossing from the "tame" ones. It would be interesting to realise these higher spin multiplets as curves on the spectral geometry.}. The BPS states are then described in this setting by open topological string amplitudes with boundaries on those curves.
The very structure of the discrete flow suggests to expand the $\tau$ functions as grand canonical partition functions for the relevant brane amplitudes. Specifically, we propose that 
\begin{equation}
\label{topo}
\tau_{\{m_i\}} (s_i,Q_i) = \sum_{n_i} s_i^{n_i} Z_{top} (q^{m_in_i} Q_i)  
\end{equation}
where $q=e^{\hbar}$, $\hbar=g_s$ being the topological string coupling, $Q_i$ the Calabi-Yau moduli and $s_i$ the fugacities for the branes amplitudes associated to BPS states with intersection numbers $m_i$ with the cycles associated to the $Q_i$
moduli. These cycles represent a basis associated to the BPS state content of theory in the relevant chamber, the intersection numbers representing the Dirac pairing among them.  It is clear from this that the expansion \eqref{topo} for the tau function 
crucially depends on the BPS chamber. Moreover, distinct flows of the BPS quivers described in the previous sections correspond to bilinear equations in distinct moduli of the Calabi-Yau. 

These bilinear equations are in the so called 
Hirota form and turn out to be equivalent to convenient combinations of blowup equations \cite{Nakajima:2005fg,Huang:2017mis}, which consist of many more equations, and suffice to determine recursively the nonperturbative part of the partition function, given the perturbative contribution \cite{Kim:2019uqw}. 

In the following we will mainly focus on the expansion of $\tau$ functions in the electric weakly coupled frame which is suitable to geometrically engineer five-dimensional gauge theories. In this case the $\tau$ function coincides with the Nekrasov-Okounkov
partition function.
\subsection{Local $\mathbb{F}_0$ and qPIII$_3$}
\label{effezero}

We first discuss the pure gauge theory case to gain some perspective. We'll then pass to the richer, and so far less understood, $N_f=2$ case. Let up point out here that the solution of PIII${_3}$ in the strong coupling expansion was worked out in \cite{Bonelli:2017gdk} in terms of the relevant Fredholm determinant (or the matrix model in the cumulants expansion). In \cite{Bonelli:2017gdk} also the relevant connection problem was solved. Since we are interested in showing the classical expansion of the cluster variables, here we discuss the different asymptotic expansion in the weak coupling $g_s\sim0$.

In our favourite example, the local $\mathbb{F}_0$,  the cluster variable $x_2=G^{-1}$, where $G$ satisfies (see \eqref{qp3})
\begin{equation}
G(qt)G(q^{-1}t)=\left(\frac{G(t)+t}{G(t)+1} \right)^2
\end{equation}
This can be written in terms of Nekrasov-Okounkov partition functions as
\begin{equation}\label{G}
G = it^{1/4}\frac{\tau_3}{\tau_1},
\end{equation}
with
\begin{align}
\tau_1=\sum_{n\in\mathbb{Z}}s^nZ(uq^n,t), && 
\tau_3=\sum_{n\in\mathbb{Z}}s^{n+\frac{1}{2}}Z\left(uq^{n+\frac{1}{2}},t \right),
\end{align}
where $Z$ is the full Nekrasov partition function for which we give explicit formulae in Appendix \ref{App:Nekrasov}. This case corresponds to the $SU(2)$ pure gauge theory with Chern-Simons level $k=0$, and we set $u_1=u_2^{-1}=u$, $q_1=q_2^{-1}=q$. The bilinear equations
 \begin{equation}
 \overline{\tau_1}\underline{\tau_1}=\tau_1^2+t^{1/2}\tau_3^2
 \end{equation}
 turn into an infinite set of equations for $Z$:
 \begin{equation}
 \sum_{n,m}s^{n+m}\left[Z(uq^n,qt)Z(uq^m,q^{-1}t)-Z(uq^n,t)Z(uq^m,t)-t^{1/2}Z(uq^{n+1/2},t)Z(uq^{m-1/2},t) \right]=0,
 \end{equation}
 where the coefficient for each power of $s$ must vanish separately. Of course, most of these equations are redundant, but everything is determined by fixing the asymptotics, i.e. the classical contribution for the partition function. Selecting the term $n+m=1$, for example, we can obtain the following equation for the $t^0$ coefficient of $Z$ (i.e. the pertirbative contribution):
 \begin{equation}
 \frac{Z_{\text{1-loop}}(uq^{-1/2})Z_\text{1-loop}(uq^{1/2})} {Z_{\text{1-loop}}^2}=\frac{1}{u^2-1}\frac{1}{u^{-2}-1},
 \end{equation} 
 which is the q-difference equation satisfied by $Z_{\text{1-loop}}$. The term $n+m=0$ allows us to determine the instanton contribution from the perturbative one in the following way:
 \begin{equation}
 \sum_nZ(uq^n;qt)Z(uq^{-n};t/q)=\sum_nZ(uq^n;t)Z(uq^{-n};t)-t^{1/2}\sum_nZ(uq^{n+1/2};t)Z(uq^{-n-1/2}).
 \end{equation}
 We can express the above equation in terms of $Z_{inst}$ as 
 \begin{gather}
 0=\sum_n t^{2n^2}u^{4n}\prod_{\e=\pm1}\left(u^{2\e}q^{2n\e};q,q^{-1}\right)_{\infty}Z_{inst}(uq^n;qt)Z_{inst}(uq^{-n};t/q) \nonumber \\
 -\sum_nt^{2n^2}\prod_{\e=\pm1}\left(u^{2\e}q^{2n\e};q,q^{-1}\right)_{\infty}Z_{inst}(uq^n;t)Z_{inst}(uq^{-n};t) \\
 +\sum_nt^{2(n+1/2)^2+1/2}\prod_{\e=\pm1}\left(u^{2\e}q^{2(n+1/2)\e};q,q^{-1}\right)_{\infty}Z_{inst}(uq^{n+1/2};t)Z_{inst}(uq^{-(n+1/2)};t)\nonumber .
 \end{gather}
 which gives a recursion relation for the coefficients of the 
 instanton expansion
 \begin{equation}
 Z_{inst}=\sum_nt^nZ_n \, .
 \end{equation}
 For example, the one-instanton term is fully determined just by the perturbative contribution:
\begin{equation}
\begin{split}
 Z_1 & =\frac{2q}{(q-1)^2}\frac{\left(u^2q;q,q^{-1} \right)_\infty\left(u^2/q;q,q^{-1} \right)_\infty\left(1/u^2q;q,q^{-1} \right)_\infty\left(q/u^2;q,q^{-1} \right)_\infty}{\left(u^2;q,q^{-1} \right)_\infty^2\left(1/u^2;q,q^{-1} \right)_\infty^2} \\
 & = \frac{2u^2q}{(q-1)^2(u^2-1)^2}\, .
\end{split}
\end{equation}
which of course correctly reproduces the one-instanton Nekrasov partition function.

It is possible to study the autonomous limit of the X-cluster variables by setting
\begin{align}
s=e^{\eta/\hbar}, && q=e^{\hbar}, && u=e^{a},
\end{align}
and sending $\hbar\rightarrow 0$. In this limit one can expand the Nekrasov partition function in the $\Omega$-background parameter as
\begin{equation}
Z(a,\hbar,t)=\exp\left\{\frac{1}{\hbar^2}\sum_{n=0}^\infty\hbar^{2n}F_n(a,t) \right\}.
\end{equation}
The behavior of the Fourier series in this limit is determined by a discrete version of the saddle point approximation \cite{Bonnet:2000dz}. Let us call $n_*$ the saddle point for $n$: for given $a,\eta$ the saddle point condition
\begin{equation}
\eta\simeq-F_0'(a+\hbar n_*)+\mathcal{O}(\hbar^2)
\end{equation}
can be satisfied only approximately. More precisely: denote by $a_*$ the "true" saddle value, given by the condition
\begin{equation}
\eta=-F_0'(a_*)+\mathcal{O}(\hbar^2).
\end{equation}
In general $a+\hbar n_*$ will be able to approximate this value only up to $\hbar$ corrections. We write this as
\begin{equation}
a+\hbar n_*=a_*+\hbar x,
\end{equation}
where the variable $x\sim O(1)$ measures the offset between the true saddle and the approximate one. We review in Appendix \ref{saddle} the computation of the leading and subleading order in the $\hbar\rightarrow 0$ limit for the dual partition functions, first performed in \cite{Bershtein:2018srt}. The result is that
\begin{equation}
\tau_1=e^{\frac{1}{\hbar^2}\left[\eta(a_*-a)+F_0(a_*,t)+F_1(a_*,t)+i\pi\tau_{SW}x^2+ \mathcal{O}(\hbar) \right]}\vartheta_3(\tau_{SW}x|\tau_{SW}),
\end{equation}
\begin{equation}
\tau_3=e^{\frac{1}{\hbar^2}\left[\eta(a_*-a)+F_0(a_*,t)+F_1(a_*,t)+i\pi\tau_{SW}x^2+ \mathcal{O}(\hbar) \right]}\vartheta_2(\tau_{SW}x|\tau_{SW}).
\end{equation}
When we consider the q-Painlev\'e transcendent, given by the cluster variable $x_2^{-1}$, the pre-factor simplifies, so that is is given by a ratio of theta functions:
\begin{equation}
G=x_2^{-1}=it^{1/4}\frac{\tau_3}{\tau_1}=it^{1/4}\frac{\vartheta_2(\tau_{SW}x|\tau_{SW})}{\vartheta_3(\tau_{SW}x|\tau_{SW})}e^{\mathcal{O}(\hbar)} .
\end{equation}
The above analysis can be easily extended to the local ${\mathbb F}_1$
case by making use of the results in \cite{Bershtein:2017swf}. 
We do not repeat it here.

\subsection{q-Painlev\'e $\text{III}_1$ and Nekrasov functions}
\label{4.2}

In the case of the gauge theory with matter, let us first focus on the 
 bilinear equations generated by the translation $T_2$. As computed in the previous section, these are 
\begin{align}
\overline{\tau_3}\tau_2=q^{1/4}t^{1/2}\overline{\tau_5}\tau_6+\tau_3\overline{\tau_2}, && \overline{\tau_6}\tau_5=\overline{\tau_5}\tau_6+q^{1/4}t^{1/2}\tau_3\overline{\tau_2}, \\
\overline{\tau_2}\underline{\tau_3}=-Q_1t^{1/2}\tau_5\tau_6+\tau_2\tau_3, && \overline{\tau_5}\underline{\tau_6}=-Q_2t^{1/2}\tau_2\tau_3+\tau_5\tau_6.
\end{align}

Let us crucially note that 
these coincide with the bilinear equations studied in \cite{Matsuhira:2018qtx}, eqs (4.5-4.8)
after relabeling
\begin{align}
(\tau_2,\tau_3)\rightarrow(\tau_1,\tau_2), && (\tau_5,\tau_6)\rightarrow(\tau_3,\tau_4),
\end{align}
and the identification
\begin{align}
Q_1=q^{-\theta_1}, && Q_2=q^{-\theta_2}.
\end{align}
In the gauge theory, 
$Q_1,Q_2$ parametrize the masses of the fundamental hypermultiplets through $\hbar\theta_i=m_i$, where $\hbar$ is the self-dual $\Omega$-background parameter, while $t$ is the instanton counting parameter. In terms of these, the time evolution is 
\begin{equation}
T_2(t)=qt,
\end{equation}
so that the discrete time evolution shifts the gauge coupling while the masses stay constant. We can therefore write the bilinear equations as q-difference equations
\begin{equation}\label{eq:BilinearqPIII1}
\begin{cases}
\tau_2(qt)\tau_3(q^{-1}t)=\tau_2(t)\tau_3(t)-Q_1t^{1/2}\tau_5(t)\tau_6(t), \\
\tau_5(qt)\tau_6(q^{-1}t)=\tau_5(t)\tau_6(t)-Q_2t^{1/2}\tau_2(t)\tau_3(t), \\
\tau_2(t)\tau_3(qt)=\tau_2(qt)\tau_3(t)+q^{1/4}t^{1/2}\tau_5(qt)\tau_6(t), \\
\tau_5(qt)\tau_6(t)=\tau_5(t)\tau_6(qt)+q^{1/4}t^{1/2}\tau_2(qt)\tau_3(t).
\end{cases}
\end{equation}
It was shown in \cite{Matsuhira:2018qtx} that the above bilinear equations are solved in terms of the dual partition function for $SU(2)$ SYM with two fundamental flavors. More precisely, in that paper it was shown that if we define
\begin{align}
Z^D_0\equiv\sum_n s^nZ(Q_1,Q_2,uq^n,t), && Z^D_{1/2}=\sum_ns^nZ(Q_1,Q_2,uq^{n+1/2},t)=Z^D_0(uq^{1/2}),
\end{align}
where $Z$ is the Nekrasov partition function for the $N_f=2$ theory, the $\tau$-functions solving \eqref{eq:BilinearqPIII1} can be written as
\begin{align}
\tau_2=Z^D_0(Q_1q^{1/2},Q_2,tq^{-1/2}), && \tau_3=Z^D_0(Q_1q^{-1/2},Q_2,tq^{1/2}), \\
\tau_5=Z^D_{1/2}(Q_1,Q_2q^{1/2},tq^{-1/2}), && \tau_6=Z^D_{1/2}(Q_1,Q_2q^{-1/2},tq^{1/2}).
\end{align}
By using also $\tau_4=T_2(\tau_2)$, $\tau_1=T_2(\tau_5) $, we can add to these
\begin{align}
\tau_1=Z^D_{1/2}(Q_1,Q_2q^{1/2},tq^{1/2}), && \tau_4=Z^D_0(Q_1q^{1/2},Q_2,tq^{1/2}).
\end{align}
Working in the same way as in Subsection \ref{effezero}, one can arrive at bilinear equations for Nekrasov functions, but differently from what happened in that simpler case, now one equation does not suffice to determine the nonperturbative contribution from the perturbative one: we have to use both the first and third equations of \eqref{eq:BilinearqPIII1}. The first equation takes the form
\begin{gather}
%\begin{equation}
%\begin{split}
\sum_nt^{2n^2}u^{2n}
Z_{\text{1-l}}Z_{\text{inst}}(Q_1q^{1/2},uq^n;tq^{1/2})Z_{\text{1-l}}Z_{\text{inst}}(Q_1q^{-1/2},uq^{-n},tq^{-1/2})=\nonumber \\
=\sum_nt^{2n^2}u^{-2n}Z_{\text{1-l}}Z_{\text{inst}}(Q_1q^{1/2},uq^n;tq^{-1/2})Z_{\text{1-l}}Z_{\text{inst}}(Q_1q^{-1/2},uq^{-n},tq^{1/2})  \\
-t^{1/2}Q_1\sum_{r\in\mathbb{Z}+1/2}t^{2r^2}u^{-2r}Z_{\text{1-l}}Z_{\text{inst}}(Q_2q^{1/2},uq^{r},tq^{-1/2})Z_{\text{1-l}}Z_{\text{inst}}(Q_2q^{-1/2},uq^{-r},tq^{1/2}). \nonumber 
%\end{split}
\end{gather}
%\end{equation}
This leads to the following equation on the one-instanton contribution:
\begin{equation}
\begin{split}
& q^{-1/2}(1-q)[Z_1(Q_1q^{-1/2})-Z_1(Q_1q^{1/2})]=\\
& =\frac{u}{Q_1}\frac{Z_{\text{1-l}}(Q_2q^{-1/2},uq^{1/2})Z_{\text{1-l}}(Q_2q^{1/2},uq^{-1/2})}{Z_{\text{1-l}}(Q_1q^{1/2})Z_{\text{1-l}}(Q_1q^{-1/2})}\\
&+\frac{1}{Q_1u}\frac{Z_{\text{1-l}}(Q_2q^{-1/2},uq^{-1/2})Z_{\text{1-l}}(Q_2q^{1/2},uq^{1/2})}{Z_{\text{1-l}}(Q_1q^{1/2})Z_{\text{1-l}}(Q_1q^{-1/2})} .
\end{split}
\end{equation}
Differently from what happened in the pure gauge case, one equation is no longer enough to determine the partition function, because we get two occurrences of the function with different shifts on the mass parameter $Q_1$.
We have to use the third equation of \eqref{eq:BilinearqPIII1}, that leads to
and the third equation, which becomes
%\begin{equation}
%\begin{split}
%&
\begin{gather}
\sum_nt^{2n^2}u^{4n}q^{n^2}Z_{\text{1-l}}Z_{\text{inst}}(Q_1q^{1/2},uq^n;tq^{-1/2})Z_{\text{1-l}}Z_{\text{inst}}(Q_1q^{-1/2},uq^{-n};tq^{3/2}) \nonumber \\ 
 =\sum_n t^{2n^2}q^{n^2}Z_{\text{1-l}}Z_{\text{inst}}(Q_1q^{1/2},uq^n;tq^{1/2})Z_{\text{1-l}}Z_{\text{inst}}(Q_1q^{-1/2},uq^{-n};tq^{1/2})\\
+q^{1/4}t^{1/2}\sum_rt^{2r^2}q^{r^2}Z_{\text{1-l}}Z_{\text{inst}}(Q_2q^{1/2},uq^r,tq^{1/2})Z_{\text{1-l}}Z_{\text{inst}}(Q_2q^{-1/2},uq^{-r};tq^{1/2})\nonumber
\end{gather}
%\end{split}
%\end{equation}
that gives an equation for the one-instanton contribution
%\begin{equation}
%\begin{split}
\begin{equation}
\begin{split}
&(1-q)\left[qZ_1(Qq^{1/2})-Z_1(Q_1q^{-1/2}) \right] \\
&=q\frac{Z_{\text{1-l}}(uq^{-1/2},Q_2q^{1/2})Z_{\text{1-l}}(uq^{1/2},Q_2q^{-1/2})}{Z_{\text{1-l}}(Q_1q^{-1/2})Z_{\text{1-l}}(Q_1q^{1/2})}\\
&+q\frac{Z_{\text{1-l}}(uq^{-1/2},Q_2q^{-1/2})Z_{\text{1-l}}(uq^{1/2},Q_2q^{1/2})}{Z_{\text{1-l}}(Q_1q^{-1/2})Z_{\text{1-l}}(Q_1q^{1/2})} .
\end{split}
\end{equation}%\end{split}
%\end{equation}
Putting the two equations together, and using the identities \eqref{eq:PochhId1} and \eqref{eq:PochhId2}, we obtain the correct one-instanton contribution
\begin{equation}
Z_1=\frac{qu^2}{(1-u^2)^2(1-q)^2}\left[\left(1-\frac{u}{Q_1} \right)\left(1-\frac{u}{Q_2} \right)+\left(1-\frac{1}{uQ_1} \right)\left(1-\frac{1}{uQ_2} \right) \right],
\end{equation}
matching the one computed by instanton counting. One can go on and compute the higher instanton contributions in an analogous way.
These two equations are enough to determine the nonperturbative contribution order by order in $t$, starting from the knowledge of the perturbative contribution, which is the $t^0$ term.

Let us finally note that all these bilinear equations could be written as lattice equations on $Q((A_2+A_1)^{(1)})$ by noting that all the various tau functions can be obtained starting from a single one, let us say $\tau_1$, since we have
\begin{align}
\tau_2=T_2^{-1}\left(T_4(\tau_1) \right), && \tau_3=T_1(\tau_1), && \tau_4=T_4(\tau_1),
\end{align}
\begin{align}
\tau_5=T_2^{-1}(\tau_1), && \tau_6=T_4^{-1}\left(T_1(\tau_1) \right),
\end{align}
so that it is possible to introduce, following \cite{doi:10.1063/1.4931481}, the tau lattice
\begin{equation}\label{fam}
\begin{split}
\tau^{k,m}_N&\equiv T_1^kT_2^mT_4^N(\tau_1)\\
& = Z^D_0\left(q^{-\frac{N+k}{2}}Q_1,q^{\frac{N-k+1}{2}}Q_2,q^{\frac{k+N}{2}}u,q^mt \right)\, .
\end{split}
\end{equation}
In this notation the original tau-variables can be denoted by
\begin{align}
\tau_1\equiv\tau^{0,0}_0, && \tau_2\equiv\tau_1^{0,-1}, && \tau_3\equiv\tau^{1,0}_0, \\
\tau_4\equiv\tau^{0,0}_1, && \tau_5\equiv\tau^{0,-1}_0, && \tau_6\equiv\tau^{1,0}_{-1},
\end{align}
and the time flows are integer shifts of the indices of the tau function \eqref{fam}. However, not all the flows are compatible with the instanton expansion: from \eqref{eq:T1qShifts} and \eqref{eq:T4qShifts} we see that the natural expansion parameter for the solution of the $T_1$ and $T_4$ flows are respectively $Q_1Q_2$ and $Q_2/Q_1$. 

In fact, the usual Nekrasov expansion, as defined in Appendix \ref{App:Nekrasov} by a converging expansion in $t$, can only solve the equations for $T_2$, which have $t$ as time parameter: this is because if we try to solve the other equations iteratively by starting with the perturbative contribution as defined in equation \eqref{eq:Nek1loop}, there is no region in parameter space where all the multiple q-Pocchammer functions with shifted arguments entering the bilinear equations have converging expressions simultaneously. To find a solution one should find an analogue of the perturbative partition function \eqref{eq:Nek1loop} which is of order zero, not in $t$, but rather in the appropriate time parameter, solving the order zero of the bilinear equations. This indeed corresponds to an expansion of the topological string partition function \eqref{topo}
in the corresponding patch in the
moduli space  
in the Topological Vertex formalism \cite{Aganagic:2003db}.

%or equivalently $\theta_i\rightarrow\theta_i+1/2$. However, this is not all. 
A preliminary analysis shows that, on top of the evolution in the mass parameters, comparing with the solution in terms of Nekrasov functions, we see that consistency requires also that
\begin{equation}\label{eq:ShiftsT1T4}
T_1(u)=q^{1/2}u
\quad
{\rm  and}
\quad
T_4(u)=q^{1/2}u.
\end{equation}
To see why this must hold, one has to consider tau functions related by time evolutions of the flows $T_1,T_4$. For example the action on the flow $T_1$ on the solutions $\tau_1,\tau_3$ (the same considerations would hold by considering the other tau functions):
\begin{align}
\tau_3=T_1(\tau_1), && \tau_1=Z^D_0(Q_1,Q_2q^{1/2},uq^{1/2},tq^{1/2}), && \tau_3=Z^D_0(Q_1q^{-1/2},Q_2,u,tq^{1/2}).
\end{align}
Equation \eqref{eq:ShiftsT1T4} is consistent with the interpretation of the flows $T_1,T_3,T_4$ as the B\"acklund transformations of $T_2$.

\section{Degeneration of cluster algebras and four-dimensional gauge theory}\label{Sec:Degeneration}

In the previous sections we saw how the cluster algebra structure on the one hand yields the q-difference equations satisfied by the partition function of the theory. On the other hand, since it conjecturally encodes the wall-crossing of states for the four-dimensional KK theory, it allows, through a generalized mutation algorithm, to produce the spectrum of the theory in a weak-coupling chamber. Further, it was observed in \cite{Closset:2019juk}, that the BPS quivers describing the purely four-dimensional theory (with all KK modes decoupled) are contained in the five-dimensional one as subquivers with two fewer nodes: roughly, one of the additional nodes is the five-dimensional instanton monopole, while the other corresponds to the KK tower of states. From the point of view of cluster integrable systems and q-Painlev\'e equations this was already realized in \cite{Bershtein:2017swf}. Graphically, to go from the 5d theory to the 4d one, one "pops" two nodes of the quiver.

We now show how it is possible to explicitly implement the operation of deleting the two nodes, that brings the five-dimensional quiver to the four-dimensional one, at the level of the full cluster algebra, so that we recover the four-dimensional description of the BPS states.
From the gauge theory point of view, the four-dimensional limit $\mathbb{R}^4\times S^1_R\rightarrow \mathbb{R}^4 $ is obtained by taking the radius of the five-dimensional circle $R\rightarrow 0$. More precisely, one has to scale the K\"ahler parameters in such a way that the KK modes and instanton particles decouple from the BPS spectrum. This limit is usually achieved by implementing the geometric engineering limit \cite{Katz:1996fh,Hollowood:2003cv}, and takes the form
\begin{align}
t=\left(\frac{R\Lambda}{2} \right)^{4-N_f}, && q=e^{-R g_s}, && u=e^{-2aR}, && R\rightarrow0.
\end{align}
We see that this limit amounts to sending 
\begin{align}
q\rightarrow 1, && t\rightarrow 0, && \frac{t}{(\log q)^{4-N_f}}\text{ finite}.
\end{align}
Because this limit involves
\begin{equation}
\log q=\log\left(\prod_i x_i \right),
\end{equation}
while the other Casimir is still given by a product of cluster variables, we can already see that it is unlikely for this limit to be able to reproduce cluster algebra transformations. Another way to see this is the case, consider the relation between X- and A-cluster variables for the case of local $\mathbb{F}_0$:
\begin{align}
x_1=\left(\frac{\tau_4}{\tau_2} \right)^2(qt)^{1/2}, && x_2=\left(\frac{\tau_1}{\tau_3} \right)^2t^{-1/2}, \\
x_3=\left(\frac{\tau_2}{\tau_4} \right)^2(qt)^{1/2}, && x_4=\left(\frac{\tau_3}{\tau_1} \right)^2t^{-1/2}.
\end{align}
Because of this, if we implement the limit $t\rightarrow 0,q\rightarrow 1$ by using the geometric engineering prescription, the tau functions, which are given in terms of five-dimensional Nekrasov partition function, will simply go to their four-dimensional limit. Then no X-cluster variable has an interesting limit. In fact, this is instead the continuous limit of the corresponding Painlev\'e equation, in which the q-discrete equations become differential equations (see e.g. \cite{Bershtein:2016aef} for the explicit implementation of the limit on the 5d Nekrasov functions).

We will now show how to instead implement the limit $q\rightarrow 1$, $t\rightarrow 0$ for the cases we considered in this paper: local $\mathbb{F}_0$, $\mathbb{F}_1$, and $\text{dP}_3$ (respectively 5d pure gauge theory without and with Chern-Simons term, and the theory with $N_f=2$ hypermultiplets), in such a way that the cluster algebra structure of the quiver is preserved: in particular we will see that:
\begin{itemize}
\item The mutations of the five-dimensional quiver degenerate to those of the four-dimensional one in terms of the reduced set of variables;
\item The q-Painlev\'e time flows (or a sub-flow, in the case of $\mathbb{F}_1$), which were given by automorphisms of the five-dimensional quivers, degenerate to appropriate sequences of mutations and permutations which are automorphisms of the four-dimensional ones.
\end{itemize}

Of course, the recipe taken to implement these limit is quite general, and we have no reason to expect it not to work for the other cases. We will implement the limit on the X-cluster variables, because they carry no ambiguity related to the choice of coefficient/extended adjacency matrix. The four-dimensional cluster A-variables can then be obtained from the X-cluster variables using the adjacency matrix as usual. However, because we are implementing this limit on the X-cluster variables, we do not have an explicit expression in terms of Nekrasov functions for the limiting system.

\subsection{From local $\mathbb{F}_0$ to the Kronecker quiver}
Recall the expression for the Casimirs in terms of the cluster variables:
\begin{align}
t=x_2^{-1}x_4^{-1}=x_1x_3/q, && q=x_1x_2x_3x_4.
\end{align}
Let us say that we want to decouple the nodes 3,4 on the corresponding quiver, so that we remain with the Kronecker quiver with nodes 1,2 (the red quiver in the Figure \ref{Fig:SubQuiverF0}). We need then to implement the limit
\begin{align}
t\rightarrow0, && q\rightarrow 1, && x_1,x_2\text{ finite},
\end{align}
which, as we argued above, is different from the geometric engineering limit. We then have to take
\begin{align}
x_3=qt/x_1\rightarrow 0, && x_4=\frac{1}{x_2 t}\rightarrow\infty, && x_3x_4\text{ finite}.
\end{align}
We are interested in the expressions for the mutations at the remaining nodes after decoupling, as well as for the q-Painlev\'e translation. For the mutations this case is very simple: the limit takes the form
\begin{align}
\mu_1(\textbf{x})= \left( \begin{array}{cccc}
x_1^{-1}, & \frac{x_2}{(1+x_1^{-1})^2}, & x_3, & (1+x_1)^2x_4
\end{array} \right)\rightarrow \left( \begin{array}{cccc}
x_1^{-1}, & \frac{x_2}{(1+x_1^{-1})^2}, & 0, & \infty
\end{array} \right),
\end{align}
\begin{equation}
\mu_2(\textbf{x})=\left( \begin{array}{cccc}
x_1(1+x_2)^2, & x_2^{-1}, & \frac{x_3}{(1+x_2^{-1})^2}, & x_4
\end{array} \right)\rightarrow \left( \begin{array}{cccc}
x_1(1+x_2)^2, & x_2^{-1}, & 0, & \infty
\end{array} \right).
\end{equation}
We see that the mutations for $x_1,x_2$ do not involve the variables $x_3,x_4$, so that no limit is actually necessary, and in fact they are already in the form of mutations for the Kronecker subquiver. Further, these mutations preserve the limiting value of $x_3,x_4$. In fact, the choice of the subquiver is completely arbitrary: by this limiting procedure we can consider any of the Kronecker subquivers of the quiver \ref{Fig:QuiverF0}. The limit is less trivial on the q-Painlev\'e flow:
\begin{equation}
T_{\mathbb{F}_0}(\textbf{x})=\left( \begin{array}{cccc}
x_2\frac{(1+x_3)^2}{(1+x_1^{-1})^2}, & x_1^{-1}, & x_4\frac{(1+x_1)^2}{(1+x_3^{-1})^2}, & x_3^{-1}
\end{array} \right)\rightarrow \left( \begin{array}{cccc}
\frac{x_2}{(1+x_1)^2}, & x_1^{-1}, 0, & \infty
\end{array} \right).
\end{equation}
Again the limiting value of $x_3,x_4$ is preserved, while in terms of operations of the Kronecker quiver the q-Painlev\'e flow becomes
\begin{equation}
T_{\textbf{F}_0}=(1,2)\mu_1,
\end{equation}
which is an automorphism of the Kronecker quiver.

\subsection{Local $\mathbb{F}_1$}

We can proceed and take the analogous limit for local $\mathbb{F}_1$, for which
\begin{align}
t=x_1x_2^{-1}x_3^2, && q=x_1x_2x_3x_4.
\end{align}
We again focus on the Kronecker subquiver with nodes 1,2, and set
\begin{align}
x_3=\left(tx_1^{-1}x_2 \right)^{1/2}\rightarrow 0, && x_4=qt^{-1/2}x_1^{-1/2}x_2^{-3/2}\rightarrow\infty.
\end{align}
The limiting behavior of the mutations is now
\begin{align}
\mu_1(\textbf{x})= \left( \begin{array}{cccc}
x_1^{-1}, & \frac{x_2}{(1+x_1^{-1})^2}, & \frac{x_3}{1+x_1^{-1}}, & (1+x_1)^3x_4
\end{array} \right)\rightarrow \left( \begin{array}{cccc}
x_1^{-1}, & \frac{x_2}{(1+x_1^{-1})^2}, & 0, & \infty
\end{array} \right),
\end{align}
\begin{equation}
\mu_2(\textbf{x})=\left( \begin{array}{cccc}
x_1(1+x_2)^2, & x_2^{-1}, & \frac{x_3}{1+x_2^{-1}}, & \frac{x_4}{1+x_2^{-1}}
\end{array} \right)\rightarrow \left( \begin{array}{cccc}
x_1(1+x_2)^2, & x_2^{-1}, & 0, & \infty
\end{array} \right),
\end{equation}
which again yields the correct limiting behavior. The q-Painlev\'e flow does not have a good limiting behavior: however its square does, since
\begin{equation}
T_{\mathbb{F}_1}^2=T_{\mathbb{F}_0}\rightarrow (1,2)\mu_1
\end{equation}
as we saw above.

\subsection{Local $\text{dP}_3 $}\label{dp3}

\begin{figure}[h]
\begin{center}
\begin{subfigure}{.4\textwidth}
\centering
\includegraphics[width=\textwidth]{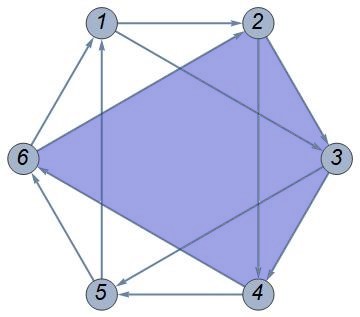}
\end{subfigure}\hfill
\begin{subfigure}{.4\textwidth}
\centering
\includegraphics[width=\textwidth]{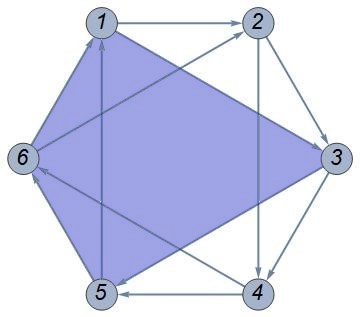}
\end{subfigure}
\caption{$N_f=2$ kite-subquivers for the discrete flow $T_2$}
\label{Fig:Nf2Sub12}
\end{center}
\end{figure}

\begin{figure}[h]
\begin{center}
\begin{subfigure}{.4\textwidth}
\centering
\includegraphics[width=\textwidth]{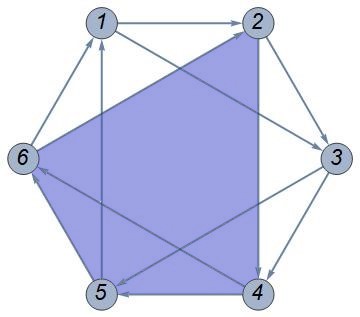}
\end{subfigure}\hfill
\begin{subfigure}{.4\textwidth}
\centering
\includegraphics[width=\textwidth]{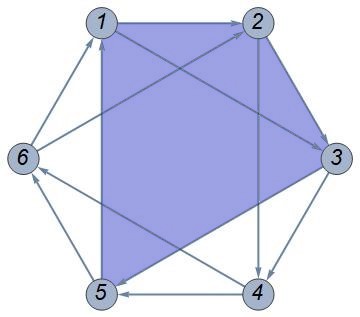}
\end{subfigure}
\caption{$N_f=2$ kite-subquivers for the discrete flow $T_1$}
\label{Fig:Nf2Sub34}
\end{center}
\end{figure}

\begin{figure}[h]
\begin{center}
\begin{subfigure}{.4\textwidth}
\centering
\includegraphics[width=\textwidth]{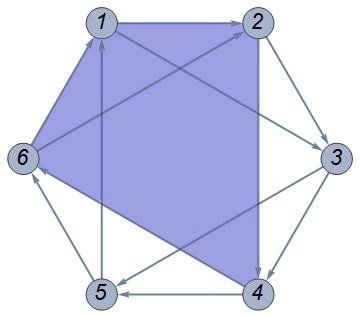}
\end{subfigure}\hfill
\begin{subfigure}{.4\textwidth}
\centering
\includegraphics[width=\textwidth]{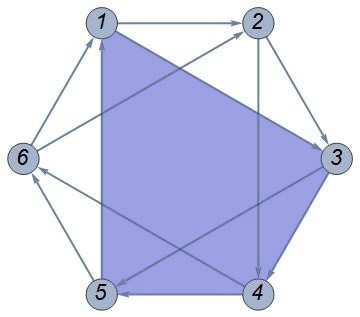}
\end{subfigure}
\caption{$N_f=2$ kite-subquivers for the discrete flow $T_3$}
\label{Fig:Nf2Sub89}
\end{center}
\end{figure}

\begin{figure}[h]
\begin{center}
\begin{subfigure}{.4\textwidth}
\centering
\includegraphics[width=\textwidth]{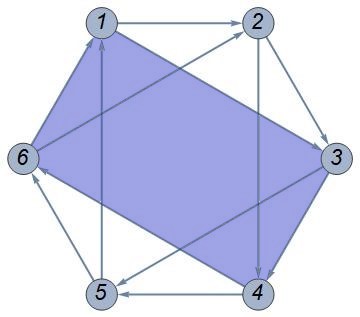}
\end{subfigure}\hfill
\begin{subfigure}{.4\textwidth}
\centering
\includegraphics[width=\textwidth]{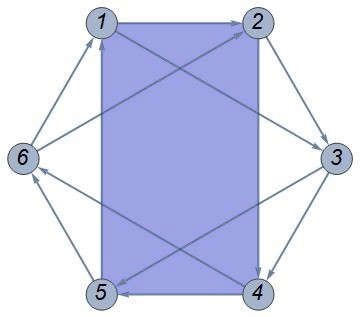}
\end{subfigure}\hfill
\begin{subfigure}{.4\textwidth}
\centering
\includegraphics[width=\textwidth]{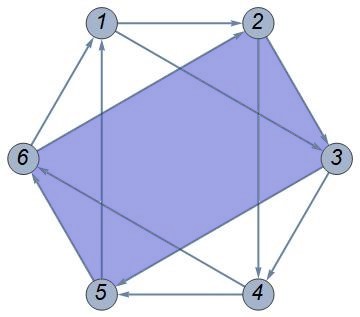}
\end{subfigure}
\caption{$N_f=2$ subquivers for the discrete flow $T_4$}
\label{Fig:Nf2Sub567}
\end{center}
\end{figure}

This case is much more interesting, because we get different decoupling limits, and only one of them is similar to the usual four-dimensional limit, involving $t\rightarrow 0$. These are related to the presence of different discrete flows. We consider as usual $T_2$ first, which we have already seen to be related to the usual weakly-couplied/instanton counting picture. In analogy to what was done in the previous cases, since
\begin{equation}
T_2(t)=qt,
\end{equation}
we take the limit
\begin{align}\label{eq:dP3LimitT2}
t\rightarrow0, && q\rightarrow 1,
\end{align}
by taking a limit on two of the cluster variables. Looking at the quiver for this case, we recognize that the subquiver with vertices 2,3,4,6 (or equivalently 1,3,5,6) gives the BPS quiver of the four-dimensional $N_f=2$ theory, as in figure \ref{Fig:Nf2Sub12}. We will focus on the former case.
Because we are "popping out" the nodes 1,5 from the quiver, we want to achieve this by implementing the limit \eqref{eq:dP3LimitT2} directly on the cluster variables. By studying the expressions for the Casimirs \eqref{eq:qPIIICas1}, \eqref{eq:qPIIICas2} we find
\begin{align}
x_3x_6=Q_1Q_2q^{1/2}, && x_1x_4=(Q_1Q_2t)^{-1}\rightarrow\infty, && x_2x_5=q^{1/2}t\rightarrow0,
\end{align}
so that we want to study the limit
\begin{align}
x_1=(x_4Q_1Q_2t)^{-1}\rightarrow\infty, && x_5=q^{1/2}tx_2^{-1}\rightarrow 0.
\end{align}
Taking the limit on the mutations we obtain
\begin{equation}
\begin{split}
\mu_2(\textbf{x}) & =\left( \begin{array}{cccccc}
x_1(1+x_2), & x_2^{-1}, & \frac{x_3}{1+x_2^{-1}}, & \frac{x_4}{1+x_2^{-1}}, & x_5, & (1+x_2)x_6
\end{array} \right)\\
& \rightarrow \left( \begin{array}{cccccc}
\infty, & x_2^{-1}, & \frac{x_3}{1+x_2^{-1}}, & \frac{x_4}{1+x_2^{-1}}, & 0, & (1+x_2)x_6
\end{array} \right),
\end{split}
\end{equation}
and similarly for the other mutations $\mu_3,\mu_4,\mu_6$, that all degenerate to the mutations of the four-dimensional quiver. The discrete flow also has a "good" limit:
\begin{align}
\begin{cases}
T_2(x_1)=\frac{1+x_5^{-1}}{(1+x_2)x_6}, \\
T_2(x_2)=\frac{x_4(1+x_5)\left(x_6(1+x_2)(1+x_5^{-1})^{-1} \right)}{(1+x_2^{-1})(1+x_3^{-1}(1+x_2^{-1})(1+x_5))}, \\
T_2(x_3)=\frac{1+x_6(1+x_2)(1+x_5^{-1})^{-1}}{x_2\left(1+x_3^{-1}(1+x_2^{-1})(1+x_5)^{-1}\right)}, \\
T_2(x_4)=\frac{1+x_2^{-1}}{x_3(1+x_5)}, \\
T_2(x_5)=\frac{x_1(1+x_2)\left(1+x_3(1+x_5)(1+x_2^{-1})^{-1} \right)}{(1+x_5^{-1})\left(1+x_6^{-1}(1+x_5^{-1})(1+x_2)^{-1} \right)}, \\
T_2(x_6)= \frac{1+x_3(1+x_5)(1+x_2^{-1})^{-1}}{x_5\left(1+x_6^{-1}(1+x_5^{-1})(1+x_2)^{-1} \right)}
\end{cases}
&& \rightarrow &&
\begin{cases}
\infty, \\
\frac{x_4}{(1+x_2^{-1})\left(1+x_3^{-1}(1+x_2^{-1})\right)}, \\
\frac{x_3}{1+x_2(1+x_3)}, \\
\frac{1+x_2^{-1}}{x_3}, \\
0, \\
x_6\left(1+x_2(1+x_3) \right).
\end{cases}
\end{align}
If we call the variables after taking the limit
\begin{align}
x_4\rightarrow X_1, && x_2\rightarrow X_2, && x_3\rightarrow X_3, && x_6\rightarrow X_4,
\end{align}
we have that the limit  of the discrete flow is
\begin{equation}
T_2^{(4d)}=(3,2,1)\mu_3\mu_2,
\end{equation}
which is an automorphism of the 4d quiver. We can  follow the same logic for the other discrete flows $T_1,T_3,T_4$: we will from now on discuss only the limits on the discrete time flows, because those on the mutations are rather simple and given by essentially the same computations as above. In the first case the flow is
\begin{align}
T_1(Q_+)=q^{-1}Q_+, && Q_+=Q_1Q_2
\end{align}
so that the natural guess for the right limit to consider is $Q_+\rightarrow \infty$, in analogy with the previous case. By looking at the Casimirs, we arrive to the conclusion that we can either decouple the nodes 1,3 or 4,6, producing the 4d $N_f=2$ subquivers in Figure \ref{Fig:Nf2Sub34}. 
The limit we want to implement is then
\begin{align}
x_4=(x_1Q_+t)^{-1}\rightarrow0, && x_6=Q_+q^{1/2}x_3^{-1}  \rightarrow \infty,
\end{align}
which gives
\begin{align}
\begin{cases}
T_1(x_1)=\frac{x_3(1+x_4)\left(1+x_5(1+x_1)(1+x_4^{-1})^{-1} \right)}{x_1\left(1+x_2^{-1}(1+x_1^{-1})(1+x_4)^{-1} \right)}, \\
T_1(x_2)=\frac{1+x_5(1+x_1)(1+x_4^{-1})^{-1}}{x_1\left(1+x_2^{-1}(1+x_1^{-1})(1+x_4)^{-1} \right)}, \\
T_1(x_3)=\frac{1+x_1^{-1}}{x_2(1+x_4)}, \\
T_1(x_4)=\frac{x_6(1+x_1)\left(1+x_2(1+x_4)(1+x_1^{-1})^{-1} \right)}{(1+x_4^{-1})\left(1+x_5^{-1}(1+x_4^{-1})(1+x_1)^{-1} \right)}, \\
T_1(x_5)=\frac{1+x_2(1+x_4)(1+x_1^{-1})^{-1}}{x_4\left(1+x_5^{-1}(1+x_4^{-1})(1+x_1)^{-1} \right)}, \\
T_1(x_6)=\frac{1+x_4^{-1}}{x_5(1+x_1)}.
\end{cases}
&& \rightarrow &&
\begin{cases}
\frac{x_3}{(1+x_1^{-1})(1+x_2^{-1}(1+x_1^{-1}))}, \\
\frac{x_2}{1+x_1(1+x_2)}, \\
\frac{1+x_1^{-1}}{x_2}, \\
0, \\
x_5\left(1+x_1(1+x_2)\right), \\
\infty.
\end{cases}
\end{align}
Which is the same 4d quiver automorphism as for $T_2$, up to permutations of the nodes. The time evolution $T_3$ is characterized by
\begin{align}
T_3(Q_+)=qQ_+, && T_3(t)=q^{-1}t,
\end{align}
so that the natural limit on the Casimirs is
\begin{align}
x_3x_6=Q_+q^{1/2}\rightarrow 0, && x_2x_5=q^{1/2}t\rightarrow\infty.
\end{align}
This can be achieved by decoupling the nodes 3,5 or 2,6, keeping the subquivers depicted in Figure \ref{Fig:Nf2Sub89}.
Choosing the former one for concreteness, we want to compute the limit
\begin{align}
x_3=Q_+q^{1/2}x_6^{-1}\rightarrow 0, && x_5=q^{1/2}tx_2^{-1}\rightarrow\infty,
\end{align}
on the discrete evolution $T_3$. This is given by
\begin{align}
\begin{cases}
T_3(x_1)=\frac{1+x_4(1+x_6)(1+x_3^{-1})^{-1}}{x_6\left(1+x_1^{-1}(1+x_6^{-1})(1+x_3) \right)}, \\
T_3(x_2)=\frac{1+x_6^{-1}}{x_1(1+x_3)}, \\
T_3(x_3)=\frac{x_5(1+x_6)\left(1+x_1(1+x_3)(1+x_6^{-1})^{-1} \right)}{(1+x_3^{-1})\left(1+x_4(1+x_3^{-1})(1+x_6)^{-1} \right)}, \\
T_3(x_4)=\frac{1+x_1(1+x_3)(1+x_6^{-1})^{-1}}{x_3\left(1+x_4^{-1}(1+x_3^{-1})(1+x_6)^{-1} \right)}, \\
T_3(x_5)=\frac{1+x_3^{-1}}{x_4(1+x_6)}, \\
T_3(x_6)=\frac{x_2(1+x_3)\left(1+x_4(1+x_6)(1+x_3^{-1})^{-1} \right)}{(1+x_6^{-1})\left(1+x_1^{-1}(1+x_6^{-1})(1+x_3)^{-1} \right)},
\end{cases}
&& \rightarrow &&
\begin{cases}
\frac{x_1}{1+x_6(1+x_1)}, \\
\frac{1+x_6^{-1}}{x_1}, \\
0, \\
x_4(1+x_6(1+x_1)), \\
\infty, \\
\frac{x_2}{(1+x_6^{-1})(1+x_1^{-1}(1+x_6^{-1}))}.
\end{cases}
\end{align}
We see that in all the cases that yielded the time evolution of q-Painlev\'e $\text{III}_1$, the degeneration of the time flow produces the same automorphism of an appropriate subquiver. It remains to study the flow $T_4$, which yielded a q-Painlev\'e IV time evolution, characterized by
\begin{align}
T_4(Q_-)=qQ_-, && Q_-=\frac{Q_2}{Q_1}.
\end{align}
In terms of the Casimirs $b_0,b_1$, this leads to
\begin{align}\label{eq:ClusterLimit4}
x_2x_4x_6=\left(qQ_- \right)^{1/2}\rightarrow 0, && x_1x_3x_5=q^{-1}Q_-^{-1/2}\rightarrow\infty.
\end{align}
To achieve this without affecting the Casimirs $a_0,a_1,a_2$ we have to decouple either the nodes 2,5, or the nodes 3,6, or the nodes 1,4, giving the subquivers in Figure \ref{Fig:Nf2Sub567}, and we will consider the first option, given by the limit
\begin{align}
x_2=(qQ_-)^{1/2}x_4^{-1}x_6^{-1}\rightarrow 0, && x_5=q^{-1}Q_-^{-1/2}x_1^{-1}x_3^{-1}\rightarrow\infty.
\end{align}

Here something similar to what happened when we studied the degeneration of the q-Painlev\'e $\text{III}_3$ associated to local $\mathbb{F}_1$ happens: recall that in that case $T_{\mathbb{F}_1}$ did not have a good degeneration limit as an automorphism of the subquiver, but rather its square did. We observed that this was related to the $\mathbb{Z}_2$-periodicity of the action of $T_{\mathbb{F}_1}$ on the BPS charges. What happens here is that not $T_4$, but rather $T_4^3$ has a good action after taking the limit, in particular only for $T_4^3$ it is true that
\begin{align}
T_4^3(x_2)\rightarrow 0, && T_4^3(x_5)\rightarrow\infty,
\end{align}
consistently with the limit. 

The resulting sub quiver is the oriented square with arrows of valency one and no diagonals with adjacency matrix
\begin{equation}
B=\left( \begin{array}{cccc}
0 & 1 & 0 & -1 \\
-1 & 0 & 1 & 0 \\
0 & -1 & 0 & 1 \\
1 & 0 & -1 & 0
\end{array} \right).
\end{equation} 

The corresponding four dimensional gauge theory has been already classified in \cite{Cecotti:2014zga} as $Q(1,1)$ and shown to correspond to $H_3$, which is the Argyres-Douglas limit of the $N_f=3$ with $SU(2)$.
All this is consistent with the reductions of the Sakai's table in Figure \ref{Fig:Sakai}.

The symmetry type of the five dimensional $SU(2)$ $N_f=2$ gauge theory is
$E_3^{(1)}$. The reduction of the $T_1$, $T_2$ and $T_3$ flows corresponds to the reduction $E_3^{(1)} \to D_2^{(1)}$, the latter being the symmetry type of the 
four dimensional $SU(2)$ $N_f=2$ gauge theory.
The reduction of the $T_4$ flow corresponds to the reduction $E_3^{(1)} \to A_2^{(1)}$, the latter being the symmetry type of the 
$H_3$ theory. 
According to Sakai's classification (see Figure \ref{Fig:Sakai}) and the analysis 
in \cite{Bonelli:2016qwg}
this flow correctly points to the Argyres-Douglas theory of $N_f=3$
which is, in the four dimensional limit, governed by the differential PIV equation.

\section{Conclusions and Outlook}

In this paper we studied the discrete flows induced by automorphisms of BPS quivers associated to Calabi-Yau geometries engineering five-dimensional quantum field theories. 
We showed that these flows provide a simple and effective way to determine the BPS spectrum of these theories, producing at the same time a set of bilinear q-difference equations
satisfied by the grand canonical partition function of topological string amplitudes. In the rank one case these are known as q-Painlev\'e equations and admit in a suitable region of the moduli space
solutions in terms of Nekrasov-Okounkov, or free fermions, partition functions. 

A very attractive feature of this approach is that  a simple symmetry principle -- the symmetry of the BPS quiver --
provides strong constraints on the BPS spectrum and contains a rich and deep set of information which goes well beyond the perturbative approaches to the same theories.
Indeed, one can show that the non-perturbative completion of topological string via a spectral determinant presentation, arising in the context of the topological string/ spectral theory correspondence, arises naturally as solution
of this system of discrete flow equations \cite{Bonelli:2017gdk}. Moreover, some of the flows associated to the BPS quiver directly link to non-perturbative phases of the corresponding gauge theory,
as we have seen for a particular flow of the local $dP_3$ geometry which describe a $(A_1,D_4)$ Argyres-Douglas point, see subsect. \ref{dp3}. A discussion of the relation between Painlev\'e equations and Argyres-Douglas points 
of four-dimensional gauge theories can be found in \cite{Bonelli:2016qwg} based on the class ${\cal S}$ description of these theories \cite{Bonelli:2011aa}.

It is also very interesting that a fully classical construction, the cluster algebra associated to the BPS quiver, contains information about the quantum geometry of the Calabi-Yau. 
Indeed the zeroes of the $\tau$-functions of the cluster algebra provide the exact spectrum of the associated quantum integrable system, as it was shown in \cite{Bonelli:2017gdk} for the local $\mathbb{F}_0$ geometry 
corresponding to relativistic Toda chain
\cite{Hatsuda:2015qzx}, and in \cite{Bonelli:2016idi,Sciarappa:2017hds} for its four-dimensional/non-relativistic limit.
Further evidence we provide in this paper is that the X-cluster 
variables of the 5d quiver flow in the 4d limit to the X-cluster variables of the corresponding four-dimensional BPS quiver, which are known to be related to the Voros symbols, i.e. exponential of the exact quantum
WKB periods of the four-dimensional integrable system \cite{Iwaki_2014}. The fact the we find that standard topological string -- or equivalently 5d gauge theory in the self-dual $\Omega$-background $\epsilon_1+\epsilon_2=0$ rather than in the Nekrasov-Shatashvili \cite{Nekrasov:2009rc}
background $\epsilon_1=0$ -- provides a quantisation of the Calabi-Yau geometry is not fully surprising from the view point of equivariant localisation. Indeed the difference between the two cases resides in a different choice of one-parameter subgroup
of the full toric action, and therefore contains the same amount of information, although under possibly very non-trivial combinatorial identities. A first instance of this phenomenon was discussed from the mathematical perspective in \cite{Nakajima:2003uh}.
For the case at hand, the non-trivial relation between the two approaches is encoded in a suitable limit of blow-up equations \cite{Grassi:2018spf,Grassi:2019coc,Gavrylenko:2020gjb,Nekrasov:2020qcq,Jeong:2020uxz}.

There are several directions to further investigate. 

Let us notice that, with respect to the framework of \cite{Alim:2011ae,Alim:2011kw},
to define a BPS chamber one should set the precise order of the arguments of the
central charges $Z(\gamma_i)$ for all the charges $\gamma_i$ in the spectrum.
While our method efficiently computes the spectrum, at least in the tame chambers, it doesn't point yet to a precise definition of the corresponding 
moduli values. This is because we still miss a link with the relevant stability conditions. Our method relies on the existence of patches in the moduli space where the topological string partition function allows finite radius converging 
expansions\footnote{See also the comments at the end of the subsection \ref{4.2} on this point.}. Let us notice that clarifying this point would prepare the skeleton of the demonstration that Kontsevich-Soibelman wall-crossing would be equivalent to the discrete equations (q-Painlev\'e and higher rank analogues) we obtain.
Moreover, the chambers we do compute are "triangular" in the sense of 
\cite{Cecotti:2015qha}. In this paper it is shown that similar chambers exist
for all the class ${\cal S}[A_1]$ theories: while multiple affinizations are generically involved, these coincide with the ones computed with our methods, at least in the examples we work out explicitly.  

We have seen that when the Calabi-Yau geometry admits several moduli there are various inequivalent flows for the same BPS quiver, but only few of them have a realisation in terms
of weakly coupled Lagrangian field theories. In some cases the other flows correspond simply to B\"acklund transformations mapping solutions one into the others. This is the case for example for the fluxes $T_1, T_2, T_3$ discussed
in subsect. \ref{Sec:dP3}. We expect that the full solution to this system of equations will be given in terms of suitable expansions of the topological 
vertex\footnote{Similar considerations appeared in \cite{Coman:2018uwk,Coman:2020qgf} for the four dimensional case.}
  \cite{Aganagic:2003db}, while its non-pertubative completion should be 
given by the spectral determinant of the corresponding $N_f=2$ spectral curve.
In other cases the flows are intrinsically non-perturbative,
like the flux $T_4$ discussed in subsect.\ref{dp3}. It would be interesting to characterise the solutions of these flows in terms of supersymmetric indices of four-dimensional gauge theories 
\cite{Cordova:2015nma, Maruyoshi:2016tqk,Maruyoshi:2016aim}.

We expect that the full refined topological string or equivalently the gauge theory in the full $\Omega$ background is captured by the {\it quantum} cluster algebra.
The bilinear equations in this case are expected to have a direct relation to the K-theoretic blow-up equations \cite{Bershtein:2018zcz}. 

We have also shown that the X-cluster variables correctly reproduce the ones of the four dimensional BPS quivers under a suitable scaling limit. It would be very interesting to further explore the relation of our results
with the ones on exact WKB methods and TBA equations \cite{Gaiotto:2009hg,Grassi:2019coc,FIORAVANTI2020135376}, possibly extending these methods to the q-difference/5d case. For the class S theories an important r\^ole
should be played by the group Hitchin system \cite{Elliott:2018yqm},
in the perspective of its quantisation \cite{Bonelli:2009zp,Bonelli:2011na}. An interesting open question is the nature of the quantum periods in five dimensions and their relation to the cluster variables appearing in the study of q-Painlev\'e equations.

In the four-dimensional case, the Painlev\'e/gauge theory correspondence \cite{Bonelli:2016qwg} extends also to non-toric cases, corresponding to isomonodromic deformation problems on higher genus Riemann surfaces,
see \cite{Bonelli:2019boe,Bonelli:2019yjd} for the genus one case. These have a 5d uplift in terms of q-Virasoro algebra \cite{Mironov:2017sqp} and matrix models
\cite{Nedelin:2015mio,Lodin:2018lbz} whose BPS quiver interpretation would be more than welcome.
Also the higher rank extension of BPS quiver flows and the associated tau-functions is to be explored in detail. As a first example, one can consider $SU(N)$ Super Yang-Mills, whose spectral determinant in matrix model
presentation was presented in \cite{Bonelli:2017ptp}. In the one period phase, this satisfies $N$-particle Toda chain equations. The corresponding cluster integrable system is discussed in \cite{Bershtein:2018srt}.
More in general, our method should extend beyond the rank $1$
case and q-Painlev\'e systems, pointing to more general results about topological string partition functions and discrete dynamical systems.

{\bf Acknowledgements}: We would like to thank
M.~Cirafici, D.~Fioravanti, M.~Semenyakin, A.~Marshakov
for interesting discussions.
Many thanks also to M.~Del Zotto and P. Gavrylenko for a careful reading of the manuscript and interesting comments.

The research of G.B. and F.DM.\ is partly supported by the INFN Iniziativa Specifica ST\&FI and by the PRIN project ``Non-perturbative Aspects Of Gauge Theories And Strings''. 
The research of  A.T.\ is partly supported by the INFN Iniziativa Specifica GAST
and by the PRIN project ``Geometria delle variet\`a algebriche''.

\begin{appendix}

\section{q-Special functions and Nekrasov functions}\label{App:Nekrasov}

The 5d instanton partition function 
\cite{Nekrasov:2003af,Bruzzo:2002xf}
for $\mathcal{N}=1$ 5d $SU(N)$ SYM with $N_f$ fundamental flavors and Chern-Simons level $k=0,1,\dots,N-1$ is given by a sum over N-tuples if partitions $\bs\lambda(\lambda_1,\dots,\lambda_N)$ with counting parameter $t$:
\begin{equation}
Z_{inst}=\sum_{\bs\lambda}t^{|\bs\lambda|} Z^{CS}_{\bs\lambda} Z_{\bs\lambda}^{fund}Z_{\bs\lambda}^{gauge}.
\end{equation}
The Chern-Simons factor given by
\begin{align}
Z^{CS}_{\bs\lambda}=\prod_{i=1}^NT_{\lambda_i}(u_i;q_1,q_2)^k, && T_\lambda(u;q_1,q_2)=\prod_{(i,j)\in\lambda}uq_1^{i-1}q_2^{j-1},
\end{align}
while the matter and gauge contributions can all be written in terms of the single building block
\begin{equation}
N_{\lambda,\mu}(u,q_1,q_2)=\prod_{s\in\lambda}\left(1-uq_2^{-a_\mu(s)-1}q_1^{l_\lambda(s)} \right)\prod_{s\in\mu}\left(1-uq_2^{a_\lambda(s)}q_1^{-l_\mu(s)-1} \right),
\end{equation}
in the following way:
\begin{equation}
Z^{fund}_{\bs\lambda}=\prod_{i=1}^{N_f}\prod_{\alpha=1}^NN_{\lambda,\emptyset}(Q_iu_\alpha),
\end{equation}
\begin{equation}
Z_{\bs\lambda}^{gauge}=\prod_{i,j=1}^N\frac{1}{N_{\lambda_i,\lambda_j}(u_i/u_j;q_1,q_2)}.
\end{equation}
The perturbative contribution is given by the following:
\begin{equation}\label{eq:Nekclass}
Z_{cl}=e^{-\log t\frac{\sum_{i=1}^N(\log u_i)^2}{2\log q_1\log q_2}-k\frac{\sum_{i=1}^N(\log u_i)^3}{6\log q_1\log q_2}},
\end{equation}
\begin{equation}\label{eq:Nek1loop}
Z_{\text{1-loop}}=\frac{\prod_{1\le \alpha\ne \beta\le N}\left(u_\alpha/u_\beta;q_1,q_2 \right)_\infty}{\prod_{i=1}^{N_f}\prod_{\alpha=1}^N\left(Q_iu_\alpha;q_1,q_2 \right)_\infty}.
\end{equation}
Here $\left(u_i/u_j;q_1,q_2 \right)_\infty$ is the multiple q-Pochhammer symbol, defined by
\begin{equation}
(z;q_1,\dots,q_n)_{\infty}\equiv\prod_{i_1,\dots,i_N=0}^{\infty}\left(1-z\prod_{k=1}^nq_k^{i_k} \right)=\exp\left(-\sum_{m=1}^\infty\frac{z^m}{m}\prod_{k=1}^N\frac{1}{1-q_k^m} \right).
\end{equation}
In all the formulae above the following notations are used, in terms of the four-dimensional gauge theory parameters:
\begin{align}
u_\alpha=e^{\beta a_\alpha}, && Q_i=e^{-\beta m_i}, && q_1=e^{\beta\epsilon_1}, && q_2=e^{\beta\epsilon_2}.
\end{align}
An important property of the double Pochhammer symbol, that has to be used repeatedly when solving the bilinear equations, is the following:
\begin{align}\label{eq:PochhId1}
\frac{\left(zq;q,q^{-1} \right)_{\infty}}{\left(z;q,q^{-1} \right)_{\infty}}=\left(zq;q\right)_{\infty}, && \frac{\left(zq^{-1};q,q^{-1} \right)_{\infty}}{\left(z;q,q^{-1} \right)_{\infty}}=\frac{1}{\left(z;q\right)_{\infty}},
\end{align}
\begin{equation}\label{eq:PochhId2}
\frac{\left(zq;q,\right)_{\infty}}{\left(z;q\right)_{\infty}}=\frac{1}{1-z}.
\end{equation}
The full partition function $Z(u;t)$ is given by $Z=Z_{cl}Z_{\rm 1-loop} Z_{inst}$.

\section{q-Painlev\'e III and IV in Tsuda's parametrization}\label{Sec:Tsuda}
We provide here the choice of parameters for the cluster algebra that reproduces the q-Painlev\`e $III_1$ and $IV$ equations of \cite{2006LMaPh..75...39T}. It turns out for this to be useful to introduce
\begin{align}
u_1=a_1\left(\frac{b_1}{b_0} \right)^{1/3}, && u_2=a_2\left(\frac{b_0}{b_1} \right)^{1/3}, && u_3=a_0\left(\frac{b_1}{b_0} \right)^{1/3}, \\
u_4=a_1\left(\frac{b_0}{b_1} \right)^{1/3}, && u_5=a_2\left(\frac{b_1}{b_0} \right)^{1/3}, && u_6=a_0\left(\frac{b_0}{b_1} \right)^{1/3},
\end{align}
Note that in \cite{2006LMaPh..75...39T} the Casimirs have a geometric meaning in terms of points blown-up on $\mathbb{P}^1\times\mathbb{P}^1$.
\subsection{qPIII}
Let us consider first the cas of qPIII. We choose as basis for the tropical semifield four independent Casimirs, and parametrize in terms of them the $y_i$'s in such a way that \eqref{eq:qPIIICas1} and \eqref{eq:qPIIICas2} are satisfied, together with the correct time evolution \eqref{eq:qPIIICasimirsT1}. We choose as independent Casimirs $a_1,a_2,q,b_0$. To match with \cite{2006LMaPh..75...39T}, we also have to make a different choice for the time evolution parameter $q$,
\begin{equation}
q=\prod_iy_i^{-1/2}.
\end{equation}
Set
\begin{align}
y_1=q^{-1/3}b_0^{2/3}a_1^{-1}, && y_2=q^{1/3}b_0^{2/3}a_2^{-1}, && y_3=q^{-4/3}b_0^{2/3}a_1a_2, \\
y_4=q^{1/3}b_0^{-2/3}a_1^{-1}, && y_5=q^{-1/3}b_0^{2/3}a_2^{-1}, &&  y_6=q^{-2/3}b_0^{-2/3}a_1a_2.
\end{align}
The time evolution is the following (we only write the relevant $\tau$-variables):
\begin{align}
\begin{cases}
\overline{\tau_2}=\tau_4, \\
\overline{\tau_3}=\frac{a_2b_0^{2/3}\tau_3\tau_4+q^{1/3}\tau_1\tau_6}{\tau_2}, \\
\overline{\tau_5}=\tau_1, \\
\overline{\tau_6}=\frac{q^{1/3}a_2\tau_1\tau_6+\tau_3\tau_4b_0^{2/3}}{\tau_5},
\end{cases}, &&
\begin{cases}
\underline{\tau_1}=\tau_5, \\
\underline{\tau_3}=\frac{a_1b_0^{2/3}\tau_5\tau_6+q^{1/3}\tau_2\tau_3}{\tau_4}, \\
\underline{\tau_4}=\tau_2, \\
\underline{\tau_6}=\frac{q^{1/3}a_1\tau_2\tau_3+b_0^{2/3}\tau_5\tau_6},
\end{cases}
\end{align}
leading to the bilinear equations
\begin{align}
\overline{\tau_3}\tau_2=q^{1/3}(\overline{\tau_5}\tau_6+u_2\tau_3\overline{\tau_2}), && \overline{\tau_6}\tau_5=b_0^{2/3}(u_5\overline{\tau_5}\tau_6+\tau_3\overline{\tau_2}), \\
\overline{\tau_2}\underline{\tau_3}=q^{1/3}(u_4\tau_5\tau_6+\tau_2\tau_3), && \overline{\tau_5}\underline{\tau_6}=b_0^{2/3}(u_1\tau_2\tau_3+\tau_5\tau_6).
\end{align}
These differ from the bilinear equations of \cite{2006LMaPh..75...39T} by different overall factors of the RHS, so in principle it would seem that they are different bilinear equations. However they are still equivalent to qPIII. If we define
\begin{align}
f=\frac{\tau_5\tau_6}{\tau_2\tau_3}, && g=\frac{\tau_1\tau_6}{\tau_3\tau_4},
\end{align}
we get the system of first-order q-difference equations
\begin{align}\label{qPIII1Transc}
\overline{f}f=a_2g\frac{g+u_5^{-1}}{g+u_2}, && \overline{g}g=\frac{f}{a_1}\frac{f+u_1}{f+u_4^{-1}},
\end{align}
which is the qPIII equation appearing in \cite{2006LMaPh..75...39T}.

\subsection{qPIV}\label{app:qPIV}

The action of $T_4$ on the tau function is the following:
\begin{align}\label{eq:TqPIV1}
\begin{cases}
\overline{\tau_1}=\tau_4, \\
\overline{\tau_2}=\frac{q^{4/3}b_0^{2/3}\tau_4\tau_5\tau_6+a_1q^{5/3}\tau_2\tau_3\tau_4+a_1a_2b_0^{4/3}\tau_1\tau_2\tau_6}{\tau_1\tau_3}, \\
\overline{\tau_3}=\tau_6, \\
\overline{\tau_4}=\frac{q^{5/3}\tau_4\tau_5\tau_6+a_1b_0^{4/3}\tau_2\tau_3\tau_4+a_1a_2q^{1/3}b_0^{2/3}\tau_1\tau_2\tau_6}{\tau_3\tau_5}, \\
\overline{\tau_5}=\tau_2, \\
\overline{\tau_6}=\frac{b_0^{4/3}\tau_4\tau_5\tau_6+q^{1/3}a_1b_0^{2/3}\tau_2\tau_3\tau_4+q^{2/3}a_1a_2\tau_1\tau_2\tau_6}{\tau_1\tau_5},
\end{cases}
\end{align}
\begin{equation}\label{eq:TqPIV2}
\begin{cases}
\underline{\tau_1}=\frac{q^{2/3}b_0^{4/3}\tau_1\tau_2\tau_3+q^{1/3}a_1\tau_1\tau_5\tau_6+a_1a_2b_0^{2/3}\tau_3\tau_4\tau_5}{\tau_2\tau_6}, \\
\underline{\tau_2}=\tau_5, \\
\underline{\tau_3}=\frac{q^{2/3}\tau_1\tau_2\tau_3+q^{1/3}a_1b_0^{2/3}\tau_1\tau_5\tau_6+a_1a_2b_0^{4/3}\tau_3\tau_4\tau_5}{\tau_2\tau_4}, \\
\underline{\tau_4}=\tau_1, \\
\underline{\tau_5}=\frac{q^{2/3}b_0^{2/3}\tau_1\tau_2\tau_3+q^{1/3}a_1b_0^{4/3}\tau_1\tau_5\tau_6+a_1a_2\tau_3\tau_4\tau_5}{\tau_4\tau_6}, \\
\underline{\tau_6}=\tau_3.
\end{cases}
\end{equation}
At a first glance these seem trilinear, rather than bilinear, equations. However, take linear combinations of \eqref{eq:TqPIV1} in such a way that the first term on the RHS cancels out:
\begin{equation}
\begin{cases}
\overline{\tau_6}\tau_1b_0^{-4/3}-q^{-5/3}\overline{\tau_4}\tau_3 =\frac{\tau_2\tau_3\tau_4a_1\left(q^{1/3}b_0^{-2/3}-q^{-5/3}b_0^{4/3} \right)+a_1a_2\tau_1\tau_2\tau_6\left(q^{2/3}b_0^{-4/3}-q^{-4/3} \right)}{\tau_5}, \\
\overline{\tau_6}\tau_5b_0^{-2/3}-q^{-4/3}\overline{\tau}_2\tau_3=\frac{a_1a_2\tau_1\tau_2\tau_6\left(q^{2/3}b_0^{-2/3}-q^{-4/3}b_0^{4/3} \right)}{\tau_1}, \\
\overline{\tau_4}\tau_5q^{-1/3}-b_0^{-2/3}\overline{\tau_2}\tau_1 =\frac{\tau_2\tau_3\tau_4a_1q^{4/3}\left(q^{-5/3}b_0^{4/3}-q^{1/3}b_0^{-2/3} \right)+a_1a_2\tau_1\tau_2\tau_6\left(1-b_0^{2/3} \right)}{\tau_3} .
\end{cases}
\end{equation}
We see that the second equation is now bilinear! We can repeat this procedure to obtain three bilinear equations from \eqref{eq:TqPIV1}:
\begin{align}
\overline{\tau_6}\tau_5b_0^{-2/3}-\overline{\tau_2}\tau_3q^{-4/3}=a_1a_2\left(q^{2/3}b_0^{-2/3}-b_0^{4/3}q^{-4/3} \right)\tau_2\tau_6, \\
\overline{\tau_6}\tau_1q^{-1/3}-\overline{\tau_4}\tau_3b_0^{-2/3}=\left(b_0^{4/3}q^{-1/3}-q^{5/3}b_0^{-2/3} \right)\tau_4\tau_6, \\
\overline{\tau_4}\tau_5q^{-1/3}-\overline{\tau_2}\tau_1b_0^{-2/3}=a_1\left(b_0^{4/3}q^{-1/3}-q^{5/3}b_0^{-2/3} \right)\tau_2\tau_4. \\
\end{align}
We can make these three second-order equations in three variables by using the \eqref{eq:TqPIV2}:
\begin{align}
\overline{\tau_6}\underline{\tau_2}b_0^{-2/3}-\overline{\tau_2}\underline{\tau_6}q^{-4/3}=a_1a_2\left(q^{2/3}b_0^{-2/3}-b_0^{4/3}q^{-4/3} \right)\tau_2\tau_6, \\
\overline{\tau_6}\underline{\tau_4}q^{-1/3}-\overline{\tau_4}\underline{\tau_6}b_0^{-2/3}=\left(b_0^{4/3}q^{-1/3}-q^{5/3}b_0^{-2/3} \right)\tau_4\tau_6, \\
\overline{\tau_4}\underline{\tau_2}q^{-1/3}-\overline{\tau_2}\underline{\tau_4}b_0^{-2/3}=a_1\left(b_0^{4/3}q^{-1/3}-q^{5/3}b_0^{-2/3} \right)\tau_2\tau_4, \\
\end{align}

These are the bilinear equations for qPIV, with the parametrization of \cite{2006LMaPh..75...39T}.

\section{q-Painlev\'e transcendents and autonomous limit}\label{saddle}
Recall from Section \ref{effezero} that the solution $G(t;q,u,s)$ to the q-Painlev\'e $\text{III}_3$ equation
\begin{equation}
G(qt)G(q^{-1}t)=\left(\frac{G(t)+t}{G(t)+1} \right)^2
\end{equation}
was expressed in terms of cluster X-variables as $G=x_2^{-1}$. In terms of dual Nekrasov functions, this translates to
\begin{equation}
G = it^{1/4}\frac{Z^D_{1/2}}{Z^D_{0}}.
\end{equation}
The dual partition functions here are defined as
\begin{align}
Z_0^D=\sum_{n\in\mathbb{Z}}s^nZ(uq^n,t), && Z_{1/2}^D=\sum_{n\in\mathbb{Z}}s^{n+\frac{1}{2}}Z\left(uq^{n+\frac{1}{2}},t \right),
\end{align}
where $Z$ is the partition function of pure five-dimensional $SU(2)$ super Yang-Mills. It is possible to study the autonomous limit of this object, following the discussion of \cite{Bershtein:2018srt} for the tau functions. One sets
\begin{align}
s=e^{\eta/\hbar}, && q=e^{\hbar}, && u=e^{a},
\end{align}
sending $\hbar\rightarrow 0$. In this limit one can write the partition function as
\begin{equation}
Z(a,\hbar,t)=\exp\left\{\frac{1}{\hbar^2}\sum_{n=0}^\infty\hbar^{2n}F_n(a,t) \right\},
\end{equation}
so that we want to take the "semi-classical" (actually autonomous) $\hbar\rightarrow 0$ limit of the dual partition functions, keeping the leading and first subleading term. Let us review the argument: in \cite{Nekrasov:2003rj} it was shown that the saddle point  of $Z$ is the Seiberg-Witten A-period: let us denote this by $a_*$. It is such that
\begin{equation}
\eta=-F_0'(a_*)+\mathcal{O}(\hbar^2).
\end{equation}
The sum over $n$ will be dominated by terms that are close to this saddle point: let us denote by $n_*$ the value of $n$ such that $a+\hbar n$ is closest to $a_*$: in other words, set
\begin{align}
a+\hbar n_*\equiv a_*+\hbar x, && x\sim \mathcal{O}(1).
\end{align}
To expand $Z_0^D$ around such value of $n$, we set $n=n_*+k$ and we sum over $k$. We also invert the relation for $n_*$,
\begin{equation}
n_*=\frac{a_*-a}{\hbar}+x
\end{equation}
in order to parametrize everything in terms of the UV Cartan parameter $a$ and the IR A-period $a_*$. The resulting saddle point expansion for $Z_0^D$ is
\begin{equation}
\begin{split}
Z_0^D & =\sum_k\exp\left\{\frac{1}{\hbar}(n_*+k)\eta+\frac{1}{\hbar^2}F_0(a_*+\hbar(k+x))+F_1(a_*+\hbar(k+x)) \right\} \\
& =\sum_k\left\{\frac{1}{\hbar^2}\left[F_0(a_*)+(a_*-a)\eta \right]+\frac{1}{\hbar}\left[(x+k)\eta+(x+k)F_0'(a_*) \right]\right\} \\
& \times\exp\left\{\frac{1}{2}(x+k)^2F_0''(a_*)+F_1(a_*) \right\} \\
& =e^{\frac{1}{\hbar^2}\left[(a_*-a)\eta+F_0(a_*)\right]+F_1(a_*)+i\pi\tau_{SW}x^2}\vartheta_3(\tau_{SW}x|\tau_{SW})e^{\mathcal{O}(\hbar)},
\end{split}
\end{equation}
where we defined the IR coupling constant
\begin{equation}
2\pi i\tau_{SW}=F_0''(a_*).
\end{equation}
An analogous computation can be done for the half-integer Fourier series:
\begin{equation}
\begin{split}
Z_{1/2}^D & =\sum_k\exp\left\{\frac{1}{\hbar}(n_*+k+1/2)\eta+\frac{1}{\hbar^2}F_0(a_*+\hbar(k+1/2+x))+F_1(a_*+\hbar(k+1/2+x)) \right\} \\
& =\sum_k\left\{\frac{1}{\hbar^2}\left[F_0(a_*)+(a_*-a)\eta \right]+\frac{1}{\hbar}\left[(x+k+1/2)\eta+(x+k+1/2)F_0'(a_*) \right]\right\} \\
& \times\exp\left\{\frac{1}{2}(x+k+1/2)^2F_0''(a_*)+F_1(a_*) \right\} \\
& =e^{\frac{1}{\hbar^2}\left[(a_*-a)\eta+F_0(a_*)\right]+F_1(a_*)+i\pi\tau_{SW}x^2}\vartheta_2(\tau_{SW}x|\tau_{SW})e^{\mathcal{O}(\hbar)} .
\end{split}
\end{equation}
When we take the ratio of these two expressions the overall factors simplify, giving
\begin{equation}
G=x_2^{-1}=it^{1/4}\frac{\vartheta_2(\tau_{SW}x|\tau_{SW})}{\vartheta_3(\tau_{SW}x|\tau_{SW})}e^{\mathcal{O}(\hbar)}.
\end{equation}
\end{appendix}
\bibliographystyle{JHEP}
\bibliography{Biblio.bib}

\providecommand{\href}[2]{#2}\begingroup\raggedright\begin{thebibliography}{10}

\bibitem{hori2003mirror}
K.~Hori, \emph{Mirror Symmetry}, Clay mathematics monographs. American
  Mathematical Society, 2003.

\bibitem{Iqbal:2012xm}
A.~Iqbal and C.~Vafa, \emph{{BPS Degeneracies and Superconformal Index in
  Diverse Dimensions}},
  \href{https://doi.org/10.1103/PhysRevD.90.105031}{\emph{Phys. Rev. D}
  {\bfseries 90} (2014) 105031}
  [\href{https://arxiv.org/abs/1210.3605}{{\ttfamily 1210.3605}}].

\bibitem{Aganagic:2003db}
M.~Aganagic, A.~Klemm, M.~Marino and C.~Vafa, \emph{{The Topological vertex}},
  \href{https://doi.org/10.1007/s00220-004-1162-z}{\emph{Commun. Math. Phys.}
  {\bfseries 254} (2005) 425}
  [\href{https://arxiv.org/abs/hep-th/0305132}{{\ttfamily hep-th/0305132}}].

\bibitem{Dijkgraaf:2007sw}
R.~Dijkgraaf, L.~Hollands, P.~Sulkowski and C.~Vafa, \emph{{Supersymmetric
  gauge theories, intersecting branes and free fermions}},
  \href{https://doi.org/10.1088/1126-6708/2008/02/106}{\emph{JHEP} {\bfseries
  02} (2008) 106} [\href{https://arxiv.org/abs/0709.4446}{{\ttfamily
  0709.4446}}].

\bibitem{2001CMaPh.220..165S}
H.~{Sakai}, \emph{{Rational Surfaces Associated with Affine Root Systemsand
  Geometry of the Painlev{\'e} Equations}},
  \href{https://doi.org/10.1007/s002200100446}{\emph{Communications in
  Mathematical Physics} {\bfseries 220} (2001) 165}.

\bibitem{Seiberg:1996bd}
N.~Seiberg, \emph{{Five-dimensional SUSY field theories, nontrivial fixed
  points and string dynamics}},
  \href{https://doi.org/10.1016/S0370-2693(96)01215-4}{\emph{Phys. Lett. B}
  {\bfseries 388} (1996) 753}
  [\href{https://arxiv.org/abs/hep-th/9608111}{{\ttfamily hep-th/9608111}}].

\bibitem{Nekrasov:2003rj}
N.~Nekrasov and A.~Okounkov, \emph{{Seiberg-Witten theory and random
  partitions}}, \href{https://doi.org/10.1007/0-8176-4467-9_15}{\emph{Prog.
  Math.} {\bfseries 244} (2006) 525}
  [\href{https://arxiv.org/abs/hep-th/0306238}{{\ttfamily hep-th/0306238}}].

\bibitem{Grassi:2014zfa}
A.~Grassi, Y.~Hatsuda and M.~Marino, \emph{{Topological Strings from Quantum
  Mechanics}}, \href{https://doi.org/10.1007/s00023-016-0479-4}{\emph{Annales
  Henri Poincare} {\bfseries 17} (2016) 3177}
  [\href{https://arxiv.org/abs/1410.3382}{{\ttfamily 1410.3382}}].

\bibitem{Bonelli:2017gdk}
G.~Bonelli, A.~Grassi and A.~Tanzini, \emph{{Quantum curves and $q$-deformed
  Painlev\'e equations}},
  \href{https://doi.org/10.1007/s11005-019-01174-y}{\emph{Lett. Math. Phys.}
  {\bfseries 109} (2019) 1961}
  [\href{https://arxiv.org/abs/1710.11603}{{\ttfamily 1710.11603}}].

\bibitem{Zamolodchikov:1994uw}
A.~B. Zamolodchikov, \emph{{Painleve III and 2-d polymers}},
  \href{https://doi.org/10.1016/0550-3213(94)90029-9}{\emph{Nucl. Phys. B}
  {\bfseries 432} (1994) 427}
  [\href{https://arxiv.org/abs/hep-th/9409108}{{\ttfamily hep-th/9409108}}].

\bibitem{Tracy:1995ax}
C.~A. Tracy and H.~Widom, \emph{{Proofs of two conjectures related to the
  thermodynamic Bethe ansatz}},
  \href{https://doi.org/10.1007/BF02100102}{\emph{Commun. Math. Phys.}
  {\bfseries 179} (1996) 667}
  [\href{https://arxiv.org/abs/solv-int/9509003}{{\ttfamily
  solv-int/9509003}}].

\bibitem{Cecotti:2011rv}
S.~Cecotti and C.~Vafa, \emph{{Classification of complete N=2 supersymmetric
  theories in 4 dimensions}}, {\emph{Surveys in differential geometry}
  {\bfseries 18} (2013) } [\href{https://arxiv.org/abs/1103.5832}{{\ttfamily
  1103.5832}}].

\bibitem{Alim:2011ae}
M.~Alim, S.~Cecotti, C.~Cordova, S.~Espahbodi, A.~Rastogi and C.~Vafa,
  \emph{{BPS Quivers and Spectra of Complete N=2 Quantum Field Theories}},
  \href{https://doi.org/10.1007/s00220-013-1789-8}{\emph{Commun. Math. Phys.}
  {\bfseries 323} (2013) 1185}
  [\href{https://arxiv.org/abs/1109.4941}{{\ttfamily 1109.4941}}].

\bibitem{Cecotti:2014zga}
S.~Cecotti and M.~Del~Zotto, \emph{{$Y$ systems, $Q$ systems, and 4D
  $\mathcal{N}=2$ supersymmetric QFT}},
  \href{https://doi.org/10.1088/1751-8113/47/47/474001}{\emph{J.\ Phys.\ A}
  {\bfseries 47} (2014) 474001}
  [\href{https://arxiv.org/abs/1403.7613}{{\ttfamily 1403.7613}}].

\bibitem{Cirafici:2017iju}
M.~Cirafici and M.~Del~Zotto, \emph{{Discrete Integrable Systems,
  Supersymmetric Quantum Mechanics, and Framed BPS States - I}},
  \href{https://arxiv.org/abs/1703.04786}{{\ttfamily 1703.04786}}.

\bibitem{Kontsevich:2008fj}
M.~Kontsevich and Y.~Soibelman, \emph{{Stability structures, motivic
  Donaldson-Thomas invariants and cluster transformations}},
  \href{https://arxiv.org/abs/0811.2435}{{\ttfamily 0811.2435}}.

\bibitem{Closset:2019juk}
C.~Closset and M.~Del~Zotto, \emph{{On 5d SCFTs and their BPS quivers. Part I:
  B-branes and brane tilings}},
  \href{https://arxiv.org/abs/1912.13502}{{\ttfamily 1912.13502}}.

\bibitem{Bershtein:2017swf}
M.~Bershtein, P.~Gavrylenko and A.~Marshakov, \emph{{Cluster integrable
  systems, $q$-Painlev\'e equations and their quantization}},
  \href{https://doi.org/10.1007/JHEP02(2018)077}{\emph{JHEP} {\bfseries 02}
  (2018) 077} [\href{https://arxiv.org/abs/1711.02063}{{\ttfamily
  1711.02063}}].

\bibitem{Bershtein:2018srt}
M.~Bershtein, P.~Gavrylenko and A.~Marshakov, \emph{{Cluster Toda chains and
  Nekrasov functions}},  \href{https://arxiv.org/abs/1804.10145}{{\ttfamily
  1804.10145}}.

\bibitem{Marshakov:2019vnz}
A.~Marshakov and M.~Semenyakin, \emph{Cluster integrable systems and spin
  chains}, \href{https://doi.org/10.1007/JHEP10(2019)100}{\emph{JHEP}
  {\bfseries 10} (2019) 100}
  [\href{https://arxiv.org/abs/1905.09921}{{\ttfamily 1905.09921}}].

\bibitem{Bershtein:2016aef}
M.~A. Bershtein and A.~I. Shchechkin, \emph{{q-deformed Painlev\'e $\tau$
  function and q-deformed conformal blocks}},
  \href{https://doi.org/10.1088/1751-8121/aa5572}{\emph{J. Phys.} {\bfseries
  A50} (2017) 085202} [\href{https://arxiv.org/abs/1608.02566}{{\ttfamily
  1608.02566}}].

\bibitem{Goncharov:2011hp}
A.~Goncharov and R.~Kenyon, \emph{{Dimers and cluster integrable systems}},
  \href{https://arxiv.org/abs/1107.5588}{{\ttfamily 1107.5588}}.

\bibitem{Yamazaki:2008bt}
M.~Yamazaki, \emph{{Brane Tilings and Their Applications}},  other thesis,
  2008.
\newblock 10.1002/prop.200810536.

\bibitem{Bershtein:2018zcz}
M.~Bershtein and A.~Shchechkin, \emph{{Painleve equations from
  Nakajima-Yoshioka blowup relations}},
  \href{https://arxiv.org/abs/1811.04050}{{\ttfamily 1811.04050}}.

\bibitem{jimbo2017cft}
M.~Jimbo, H.~Nagoya and H.~Sakai, \emph{Cft approach to the q-painlev{\'e} vi
  equation}, {\emph{Journal of Integrable Systems} {\bfseries 2} (2017) }.

\bibitem{Matsuhira:2018qtx}
Y.~Matsuhira and H.~Nagoya, \emph{{Combinatorial expressions for the tau
  functions of $q$-Painlevé V and III equations}},
  \href{https://arxiv.org/abs/1811.03285}{{\ttfamily 1811.03285}}.

\bibitem{Nagoya:2020maa}
H.~Nagoya, \emph{{On $q$-isomonodromic deformations and $q$-Nekrasov
  functions}},  \href{https://arxiv.org/abs/2004.13916}{{\ttfamily
  2004.13916}}.

\bibitem{Nakajima:2005fg}
H.~Nakajima and K.~Yoshioka, \emph{{Instanton counting on blowup. II.
  K-theoretic partition function}},
  \href{https://arxiv.org/abs/math/0505553}{{\ttfamily math/0505553}}.

\bibitem{Shchechkin:2020ryb}
A.~Shchechkin, \emph{{Blowup relations on $\mathbb{C}^2/\mathbb{Z}_2$ from
  Nakajima-Yoshioka blowup relations}},
  \href{https://arxiv.org/abs/2006.08582}{{\ttfamily 2006.08582}}.

\bibitem{2006LMaPh..75...39T}
T.~{Tsuda}, \emph{{Tau Functions of q-Painlev{\'e} III and IV Equations}},
  \href{https://doi.org/10.1007/s11005-005-0037-3}{\emph{Letters in
  Mathematical Physics} {\bfseries 75} (2006) 39}.

\bibitem{Alim:2011kw}
M.~Alim, S.~Cecotti, C.~Cordova, S.~Espahbodi, A.~Rastogi and C.~Vafa,
  \emph{{$\mathcal{N} = 2$ quantum field theories and their BPS quivers}},
  \href{https://doi.org/10.4310/ATMP.2014.v18.n1.a2}{\emph{Adv. Theor. Math.
  Phys.} {\bfseries 18} (2014) 27}
  [\href{https://arxiv.org/abs/1112.3984}{{\ttfamily 1112.3984}}].

\bibitem{Galakhov:2013oja}
D.~Galakhov, P.~Longhi, T.~Mainiero, G.~W. Moore and A.~Neitzke, \emph{{Wild
  Wall Crossing and BPS Giants}},
  \href{https://doi.org/10.1007/JHEP11(2013)046}{\emph{JHEP} {\bfseries 11}
  (2013) 046} [\href{https://arxiv.org/abs/1305.5454}{{\ttfamily 1305.5454}}].

\bibitem{Cordova:2015vma}
C.~Cordova, \emph{{Regge trajectories in $ \mathcal{N} $ = 2 supersymmetric
  Yang-Mills theory}},
  \href{https://doi.org/10.1007/JHEP09(2016)020}{\emph{JHEP} {\bfseries 09}
  (2016) 020} [\href{https://arxiv.org/abs/1502.02211}{{\ttfamily
  1502.02211}}].

\bibitem{doi:10.1063/1.4931481}
N.~Joshi, N.~Nakazono and Y.~Shi, \emph{Lattice equations arising from discrete
  painlev\'e systems. i. $(a_2 + a_1)^{(1)}$ and $(a_1+a_1')^{(1)}$ cases},
  \href{https://doi.org/10.1063/1.4931481}{\emph{Journal of Mathematical
  Physics} {\bfseries 56} (2015) 092705}
  [\href{https://arxiv.org/abs/https://doi.org/10.1063/1.4931481}{{\ttfamily
  https://doi.org/10.1063/1.4931481}}].

\bibitem{fomin2002cluster}
S.~Fomin and A.~Zelevinsky, \emph{Cluster algebras i: foundations},
  {\emph{Journal of the American Mathematical Society} {\bfseries 15} (2002)
  497}.

\bibitem{2006math......2259F}
S.~{Fomin} and A.~{Zelevinsky}, \emph{{Cluster algebras IV: Coefficients}},
  {\emph{arXiv Mathematics e-prints} (2006) math/0602259}
  [\href{https://arxiv.org/abs/math/0602259}{{\ttfamily math/0602259}}].

\bibitem{Fock:2014ifa}
V.~Fock and A.~Marshakov, \emph{{Loop groups, Clusters, Dimers and Integrable
  systems}},  \href{https://arxiv.org/abs/1401.1606}{{\ttfamily 1401.1606}}.

\bibitem{joshi2019discrete}
N.~Joshi, \emph{Discrete Painlev{\'e} Equations}, CBMS Regional Conference
  Series in Mathematics. Conference Board of the Mathematical Sciences, 2019.

\bibitem{Okamoto1979SurLF}
K.~Okamoto, \emph{Sur les feuilletages associ{\'e}s aux {\'e}quation du second
  ordre {\`a} points critiques fixes de p. painlev{\'e} espaces des conditions
  initiales},  1979.

\bibitem{2003}
T.~Takenawa, \emph{Weyl group symmetry of type $d_5^{(1)}$ in the
  $q$-painlev\'e v equation},
  \href{https://doi.org/10.1619/fesi.46.173}{\emph{Funkcialaj Ekvacioj}
  {\bfseries 46} (2003) 173}.

\bibitem{2009arXiv0910.4439K}
K.~{Kajiwara}, N.~{Nakazono} and T.~{Tsuda}, \emph{{Projective reduction of the
  discrete Painlev{\'e} system of type $(A_2+A_1)^{(1)}$}}, {\emph{arXiv
  e-prints} (2009) arXiv:0910.4439}
  [\href{https://arxiv.org/abs/0910.4439}{{\ttfamily 0910.4439}}].

\bibitem{Bonelli:2016qwg}
G.~Bonelli, O.~Lisovyy, K.~Maruyoshi, A.~Sciarappa and A.~Tanzini, \emph{{On
  Painlev\'e/gauge theory correspondence}},
  \href{https://doi.org/10.1007/s11005-017-0983-6}{\emph{Letters in
  Mathematical Physics} {\bfseries 107} (2017) 2359}
  [\href{https://arxiv.org/abs/1612.06235}{{\ttfamily 1612.06235}}].

\bibitem{Grassi:2018spf}
A.~Grassi and J.~Gu, \emph{{Argyres-Douglas theories, Painlev\'e II and quantum
  mechanics}},  \href{https://arxiv.org/abs/1803.02320}{{\ttfamily
  1803.02320}}.

\bibitem{Huang:2017mis}
M.-x. Huang, K.~Sun and X.~Wang, \emph{{Blowup Equations for Refined
  Topological Strings}},
  \href{https://doi.org/10.1007/JHEP10(2018)196}{\emph{JHEP} {\bfseries 10}
  (2018) 196} [\href{https://arxiv.org/abs/1711.09884}{{\ttfamily
  1711.09884}}].

\bibitem{Kim:2019uqw}
J.~Kim, S.-S. Kim, K.-H. Lee, K.~Lee and J.~Song, \emph{{Instantons from
  Blow-up}}, \href{https://doi.org/10.1007/JHEP11(2019)092}{\emph{JHEP}
  {\bfseries 11} (2019) 092}
  [\href{https://arxiv.org/abs/1908.11276}{{\ttfamily 1908.11276}}].

\bibitem{Bonnet:2000dz}
G.~Bonnet, F.~David and B.~Eynard, \emph{{Breakdown of universality in multicut
  matrix models}}, \href{https://doi.org/10.1088/0305-4470/33/38/307}{\emph{J.
  Phys. A} {\bfseries 33} (2000) 6739}
  [\href{https://arxiv.org/abs/cond-mat/0003324}{{\ttfamily
  cond-mat/0003324}}].

\bibitem{Katz:1996fh}
S.~H. Katz, A.~Klemm and C.~Vafa, \emph{{Geometric engineering of quantum field
  theories}}, \href{https://doi.org/10.1016/S0550-3213(97)00282-4}{\emph{Nucl.
  Phys.} {\bfseries B497} (1997) 173}
  [\href{https://arxiv.org/abs/hep-th/9609239}{{\ttfamily hep-th/9609239}}].

\bibitem{Hollowood:2003cv}
T.~J. Hollowood, A.~Iqbal and C.~Vafa, \emph{{Matrix models, geometric
  engineering and elliptic genera}},
  \href{https://doi.org/10.1088/1126-6708/2008/03/069}{\emph{JHEP} {\bfseries
  03} (2008) 069} [\href{https://arxiv.org/abs/hep-th/0310272}{{\ttfamily
  hep-th/0310272}}].

\bibitem{Bonelli:2011aa}
G.~Bonelli, K.~Maruyoshi and A.~Tanzini, \emph{{Wild Quiver Gauge Theories}},
  \href{https://doi.org/10.1007/JHEP02(2012)031}{\emph{JHEP} {\bfseries 02}
  (2012) 031} [\href{https://arxiv.org/abs/1112.1691}{{\ttfamily 1112.1691}}].

\bibitem{Hatsuda:2015qzx}
Y.~Hatsuda and M.~Marino, \emph{{Exact quantization conditions for the
  relativistic Toda lattice}},
  \href{https://doi.org/10.1007/JHEP05(2016)133}{\emph{JHEP} {\bfseries 05}
  (2016) 133} [\href{https://arxiv.org/abs/1511.02860}{{\ttfamily
  1511.02860}}].

\bibitem{Bonelli:2016idi}
G.~Bonelli, A.~Grassi and A.~Tanzini, \emph{{Seiberg Witten theory as a Fermi
  gas}}, \href{https://doi.org/10.1007/s11005-016-0893-z}{\emph{Lett. Math.
  Phys.} {\bfseries 107} (2017) 1}
  [\href{https://arxiv.org/abs/1603.01174}{{\ttfamily 1603.01174}}].

\bibitem{Sciarappa:2017hds}
A.~Sciarappa, \emph{{Exact relativistic Toda chain eigenfunctions from
  Separation of Variables and gauge theory}},
  \href{https://doi.org/10.1007/JHEP10(2017)116}{\emph{JHEP} {\bfseries 10}
  (2017) 116} [\href{https://arxiv.org/abs/1706.05142}{{\ttfamily
  1706.05142}}].

\bibitem{Iwaki_2014}
K.~Iwaki and T.~Nakanishi, \emph{Exact {WKB} analysis and cluster algebras},
  \href{https://doi.org/10.1088/1751-8113/47/47/474009}{\emph{Journal of
  Physics A: Mathematical and Theoretical} {\bfseries 47} (2014) 474009}.

\bibitem{Nekrasov:2009rc}
N.~A. Nekrasov and S.~L. Shatashvili, \emph{{Quantization of Integrable Systems
  and Four Dimensional Gauge Theories}},  in \emph{{Proceedings, 16th
  International Congress on Mathematical Physics (ICMP09): Prague, Czech
  Republic, August 3-8, 2009}}, pp.~265--289, 2009,
  \href{https://arxiv.org/abs/0908.4052}{{\ttfamily 0908.4052}},
  \href{https://doi.org/10.1142/9789814304634_0015}{DOI}.

\bibitem{Nakajima:2003uh}
H.~Nakajima and K.~Yoshioka, \emph{{Lectures on instanton counting}},  in
  \emph{{CRM Workshop on Algebraic Structures and Moduli Spaces Montreal,
  Canada, July 14-20, 2003}}, 2003,
  \href{https://arxiv.org/abs/math/0311058}{{\ttfamily math/0311058}}.

\bibitem{Grassi:2019coc}
A.~Grassi, J.~Gu and M.~Mari\~no, \emph{{Non-perturbative approaches to the
  quantum Seiberg-Witten curve}},
  \href{https://arxiv.org/abs/1908.07065}{{\ttfamily 1908.07065}}.

\bibitem{Gavrylenko:2020gjb}
P.~Gavrylenko, A.~Marshakov and A.~Stoyan, \emph{{Irregular conformal blocks,
  Painlev\'e III and the blow-up equations}},
  \href{https://arxiv.org/abs/2006.15652}{{\ttfamily 2006.15652}}.

\bibitem{Nekrasov:2020qcq}
N.~Nekrasov, \emph{{Blowups in BPS/CFT correspondence, and Painlev\'e VI}},
  \href{https://arxiv.org/abs/2007.03646}{{\ttfamily 2007.03646}}.

\bibitem{Jeong:2020uxz}
S.~Jeong and N.~Nekrasov, \emph{{Riemann-Hilbert correspondence and blown up
  surface defects}},  \href{https://arxiv.org/abs/2007.03660}{{\ttfamily
  2007.03660}}.

\bibitem{Cecotti:2015qha}
S.~Cecotti and M.~Del~Zotto, \emph{{Galois covers of $\mathcal{N}=2$ BPS
  spectra and quantum monodromy}},
  \href{https://doi.org/10.4310/ATMP.2016.v20.n6.a1}{\emph{Adv. Theor. Math.
  Phys.} {\bfseries 20} (2016) 1227}
  [\href{https://arxiv.org/abs/1503.07485}{{\ttfamily 1503.07485}}].

\bibitem{Coman:2018uwk}
I.~Coman, E.~Pomoni and J.~Teschner, \emph{{From quantum curves to topological
  string partition functions}},
  \href{https://arxiv.org/abs/1811.01978}{{\ttfamily 1811.01978}}.

\bibitem{Coman:2020qgf}
I.~Coman, P.~Longhi and J.~Teschner, \emph{{From quantum curves to topological
  string partition functions II}},
  \href{https://arxiv.org/abs/2004.04585}{{\ttfamily 2004.04585}}.

\bibitem{Cordova:2015nma}
C.~Cordova and S.-H. Shao, \emph{{Schur Indices, BPS Particles, and
  Argyres-Douglas Theories}},
  \href{https://doi.org/10.1007/JHEP01(2016)040}{\emph{JHEP} {\bfseries 01}
  (2016) 040} [\href{https://arxiv.org/abs/1506.00265}{{\ttfamily
  1506.00265}}].

\bibitem{Maruyoshi:2016tqk}
K.~Maruyoshi and J.~Song, \emph{{Enhancement of Supersymmetry via
  Renormalization Group Flow and the Superconformal Index}},
  \href{https://doi.org/10.1103/PhysRevLett.118.151602}{\emph{Phys. Rev. Lett.}
  {\bfseries 118} (2017) 151602}
  [\href{https://arxiv.org/abs/1606.05632}{{\ttfamily 1606.05632}}].

\bibitem{Maruyoshi:2016aim}
K.~Maruyoshi and J.~Song, \emph{{$ \mathcal{N}=1 $ deformations and RG flows of
  $ \mathcal{N}=2 $ SCFTs}},
  \href{https://doi.org/10.1007/JHEP02(2017)075}{\emph{JHEP} {\bfseries 02}
  (2017) 075} [\href{https://arxiv.org/abs/1607.04281}{{\ttfamily
  1607.04281}}].

\bibitem{Gaiotto:2009hg}
D.~Gaiotto, G.~W. Moore and A.~Neitzke, \emph{{Wall-crossing, Hitchin Systems,
  and the WKB Approximation}},
  \href{https://arxiv.org/abs/0907.3987}{{\ttfamily 0907.3987}}.

\bibitem{FIORAVANTI2020135376}
D.~Fioravanti and D.~Gregori, \emph{Integrability and cycles of deformed n=2
  gauge theory},
  \href{https://doi.org/https://doi.org/10.1016/j.physletb.2020.135376}{\emph{Physics
  Letters B} {\bfseries 804} (2020) 135376}.

\bibitem{Elliott:2018yqm}
C.~Elliott and V.~Pestun, \emph{{Multiplicative Hitchin Systems and
  Supersymmetric Gauge Theory}},
  \href{https://arxiv.org/abs/1812.05516}{{\ttfamily 1812.05516}}.

\bibitem{Bonelli:2009zp}
G.~Bonelli and A.~Tanzini, \emph{{Hitchin systems, N=2 gauge theories and
  W-gravity}},
  \href{https://doi.org/10.1016/j.physletb.2010.06.027}{\emph{Phys. Lett.}
  {\bfseries B691} (2010) 111}
  [\href{https://arxiv.org/abs/0909.4031}{{\ttfamily 0909.4031}}].

\bibitem{Bonelli:2011na}
G.~Bonelli, K.~Maruyoshi and A.~Tanzini, \emph{{Quantum Hitchin Systems via
  ${\beta}$ -Deformed Matrix Models}},
  \href{https://doi.org/10.1007/s00220-017-3053-0}{\emph{Commun. Math. Phys.}
  {\bfseries 358} (2018) 1041}
  [\href{https://arxiv.org/abs/1104.4016}{{\ttfamily 1104.4016}}].

\bibitem{Bonelli:2019boe}
G.~Bonelli, F.~Del~Monte, P.~Gavrylenko and A.~Tanzini,
  \emph{{$\mathcal{N}=2^*$ gauge theory, free fermions on the torus and
  Painlev\'e VI}},
  \href{https://doi.org/10.1007/s00220-020-03743-y}{\emph{Commun. Math. Phys.}
  {\bfseries 377} (2020) 1381}
  [\href{https://arxiv.org/abs/1901.10497}{{\ttfamily 1901.10497}}].

\bibitem{Bonelli:2019yjd}
G.~Bonelli, F.~Del~Monte, P.~Gavrylenko and A.~Tanzini, \emph{{Circular quiver
  gauge theories, isomonodromic deformations and $W_N$ fermions on the torus}},
   \href{https://arxiv.org/abs/1909.07990}{{\ttfamily 1909.07990}}.

\bibitem{Mironov:2017sqp}
A.~Mironov and A.~Morozov, \emph{{q-Painlev\'e equation from Virasoro
  constraints}},
  \href{https://doi.org/10.1016/j.physletb.2018.08.046}{\emph{Phys. Lett. B}
  {\bfseries 785} (2018) 207}
  [\href{https://arxiv.org/abs/1708.07479}{{\ttfamily 1708.07479}}].

\bibitem{Nedelin:2015mio}
A.~Nedelin and M.~Zabzine, \emph{{q-Virasoro constraints in matrix models}},
  \href{https://doi.org/10.1007/JHEP03(2017)098}{\emph{JHEP} {\bfseries 03}
  (2017) 098} [\href{https://arxiv.org/abs/1511.03471}{{\ttfamily
  1511.03471}}].

\bibitem{Lodin:2018lbz}
R.~Lodin, A.~Popolitov, S.~Shakirov and M.~Zabzine, \emph{{Solving q-Virasoro
  constraints}}, \href{https://doi.org/10.1007/s11005-019-01216-5}{\emph{Lett.
  Math. Phys.} {\bfseries 110} (2020) 179}
  [\href{https://arxiv.org/abs/1810.00761}{{\ttfamily 1810.00761}}].

\bibitem{Bonelli:2017ptp}
G.~Bonelli, A.~Grassi and A.~Tanzini, \emph{{New results in $\mathcal{N}=2$
  theories from non-perturbative string}},
  \href{https://doi.org/10.1007/s00023-017-0643-5}{\emph{Annales Henri
  Poincare} {\bfseries 19} (2018) 743}
  [\href{https://arxiv.org/abs/1704.01517}{{\ttfamily 1704.01517}}].

\bibitem{Nekrasov:2003af}
N.~A. Nekrasov, \emph{{Seiberg-Witten prepotential from instanton counting}},
  in \emph{{International Congress of Mathematicians (ICM 2002) Beijing, China,
  August 20-28, 2002}}, 2003,
  \href{https://arxiv.org/abs/hep-th/0306211}{{\ttfamily hep-th/0306211}}.

\bibitem{Bruzzo:2002xf}
U.~Bruzzo, F.~Fucito, J.~F. Morales and A.~Tanzini, \emph{{Multiinstanton
  calculus and equivariant cohomology}},
  \href{https://doi.org/10.1088/1126-6708/2003/05/054}{\emph{JHEP} {\bfseries
  05} (2003) 054} [\href{https://arxiv.org/abs/hep-th/0211108}{{\ttfamily
  hep-th/0211108}}].

\end{thebibliography}\endgroup

\end{document}